\documentclass[entropy,article,submit,moreauthors,pdftex]{Definitions/mdpi} 

\usepackage{natbib}

\usepackage{multirow}
\newcommand{\spc}[1]{}%\hspace{#1 em}}
\newcommand{\tb}{\\[0.0ex]}
\newcommand{\ts}{\\[0.7ex]}
\newcommand{\tn}{\\[1.5ex]}

\preto{\abstractkeywords}{\nolinenumbers}

\firstpage{1} 
\pubvolume{xx}
\issuenum{1}
\articlenumber{5}
\pubyear{2020}
\copyrightyear{2020}
\history{Received: date; Accepted: date; Published: date}
\Title{Filtering Statistics on Networks}

\Author{G. J.  Baxter $^{1}$, R. A. da Costa $^{1}$*, S.~N. Dorogovtsev $^{1}$ and J.~F.~F. Mendes$^{1}$}
\AuthorNames{G. J. Baxter, R. A. da Costa, S. N. Dorogovtsev, J. F. F. Mendes}

\address{%
$^{1}$ \quad Department of Physics, University of Aveiro de $\&$ I3N,\\ Campus Universit\'ario de Santiago, 3810-193 Aveiro, Portugal}
\corres{Correspondence: americo.costa@ua.pt}

\abstract{
We explored the statistics of filtering of simple patterns on a number of deterministic and random graphs as a tractable simple example of information processing in complex systems. 
In this problem, multiple inputs map to the same output, and the statistics of filtering is represented by the distribution of this degeneracy. 
For a few simple filter patterns on a ring we obtained an exact solution of the problem and described numerically more difficult filter setups.
For each of the filter patterns and networks we found a few numbers essentially describing the statistics of filtering and compared them for different networks. 
Our results for networks with diverse architectures appear to be essentially determined by two factors: whether the graphs structure is deterministic or random, and the vertex degree.
We find that filtering in random graphs produces a much richer statistics than in deterministic graphs. This statistical richness is reduced by increasing the graph's degree.
}


\begin{document}
%%%%%%%%%%%%%%%%%%%%%%%%%%%%%%%%%%%%%%%%%%

%%%%%%%%%%%%%%%%%%%%%%%%%%%%%%%%%%%%%%%%%%

\section{Introduction}

Filtering is the processing of an input signal to produce an output signal according to some rule, based on the content of the input. The filter does not add information, with the number of possible outputs being less than (or at most equal to) the number of possible inputs. Thus, outputs are degenerate: multiple inputs map to the same output. Even very simple filters can produce a complex distribution of degeneracies \cite{baxter2020complex}.
This characteristic, of a nontrivial mapping of a configuration space to a smaller set of final configurations, also appears in sampling, compression and more general information processing \cite{song2017emergence,baek2011zipf}, and in numerous complex systems, including the basins of attraction of local minima in spin glasses, and deep learning neural networks \cite{hartmann2005phase,mezard1987spin,keskar2016large}.
Understanding the statistics of degeneracies can give important insight into these systems. In a previous work \cite{baxter2020complex}, we showed that a simple filtering problem produces analogous behaviour of the degeneracy distribution to these more complex systems, 
and that one can obtain exact results up to large system sizes that are simply not accessible in more complex problems.   

\begin{figure}[H]
\centering
\includegraphics[width=0.55\textwidth]{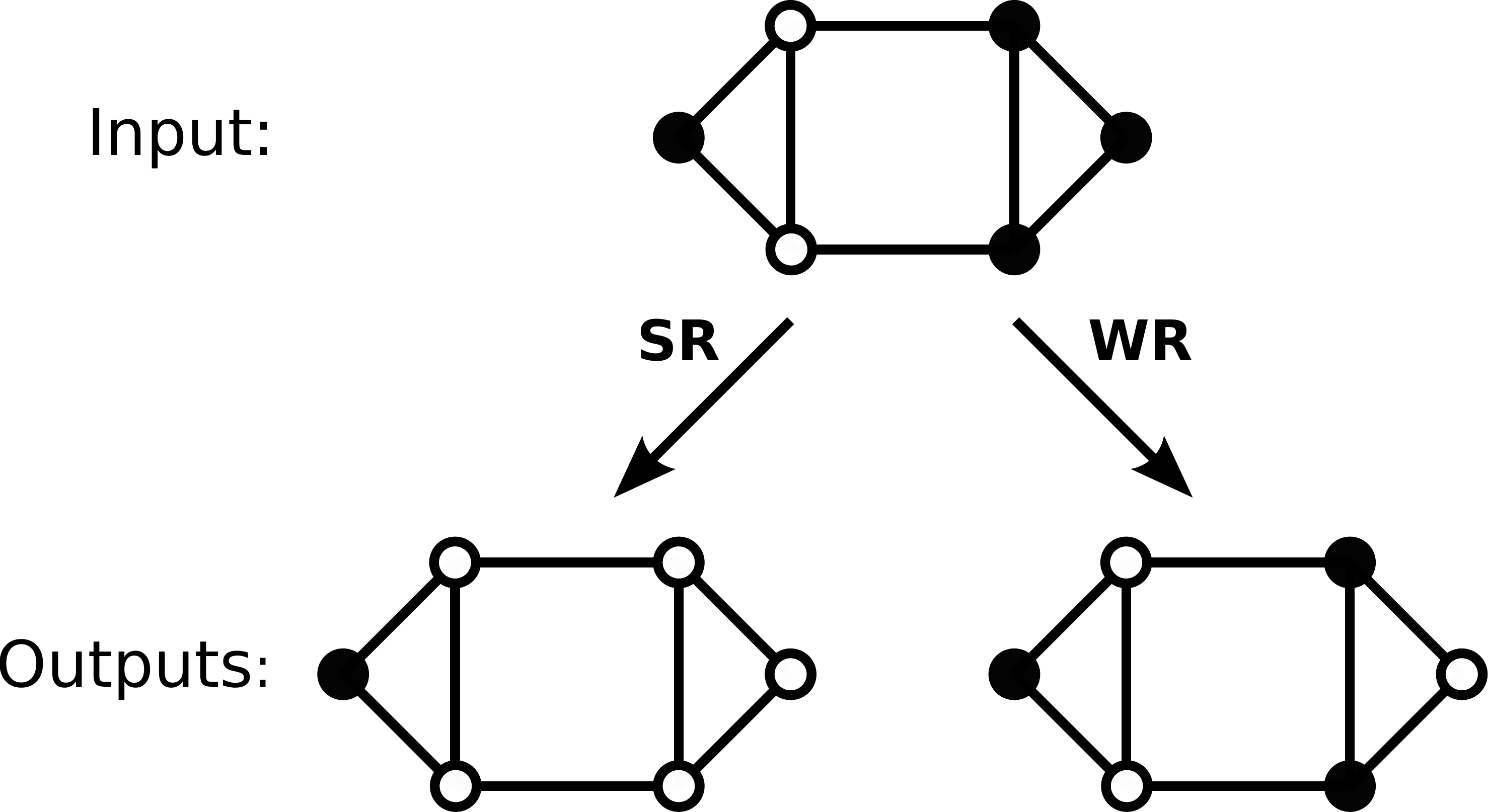}
\caption{
%Illustration of the application of the SR and WR filters on an example graph.
%The graph is the same at the input and the output. 
Application of different filters to a set of zeros and ones place on a graph.
Each node of the input and output graphs is in one of two states, namely 0 (open circles) or 1 (closed circles).
%For each filter, the state of the output nodes is determined by the state of the corresponding input node and its neighbours.
In the SR filter, an output node is one only when the corresponding input node is one and all its neighbours are zero.
In the WR filter, an output node is one when the corresponding input node is one and one or more of its neighbours are zero.
}
\label{graph_fig}
\end{figure}

Numerous studies have shown that the heterogeneous structure of interactions between elements of a complex system, usually represented as a complex network, can have a profound effect on the properties of the system~\cite{dorogovtsev2008critical}.
Here we examine a simple filtering process on a network. The input consists of the binary states of nodes in a given network. The filter outputs a $1$ for every instance of a particular pattern of states on a node and its immediate neighbours, and a $0$ when the pattern is absent. This generalises the filtering problem examined in Ref.~\cite{baxter2020complex} for binary inputs in a cyclical string (ring).
The process applied to a small graph is represented in Figure~\ref{graph_fig}. 
We studied this problem on a variety of degree-regular graphs.
We studied this problem on a variety of degree-regular graphs. 
We show that one may find the exact degeneracy distribution corresponding to the complete set of all possible inputs, up to relatively large system sizes, for any given graph. Just as in our previous study on rings, we show that the principal characteristics of the degeneracy distribution are described asymptotically by three key numbers. These numbers may be obtained exactly by simple arguments.
%

%%%=================================================================
%%%=================================================================
%%%=================================================================
%%%=================================================================

This problem serves as a tractable simple model to explore information processing in complex systems. In a graph, the connections between nodes create complex interactions between the filter output at each node. We show that the degeneracy distribution correctly captures this complexity. In particular, the entropy of the degeneracy distribution, called the {\em relevance}~\cite{cubero2018minimally} is lower in deterministically constructed graphs, and higher in random graphs. We show that relevance is maximum when the graph degree takes its smallest value greater than two.
We compared two different filters, and found that the stronger filter (detecting less easily satisfied conditions) is more informative, because it is more sensitive to the state of neighboring nodes.
Interestingly, as Figure~\ref{numbers_vs_q} demonstrates, our results for regular graphs of diverse architectures essentially depend only on a vertex degree.

%%%%%%%%%%%

\section{Results}

\subsection{Filtering statistics on a ring}
\label{s2} 

For orientation, we begin by studying nodes located on a ring.
The input is a set of $N$ strings of zeroes and ones $\{x_i\}$, $x_i=0,1$,  of length $n$, assuming the periodic condition $x_1=x_{n+1}$. 
We consider  the complete set of all possible unique inputs. Its size $N$ is determined by the size $n$ of inputs, $N = 2^n$.

\begin{figure}[H]
\centering
\includegraphics[width=0.68\textwidth]{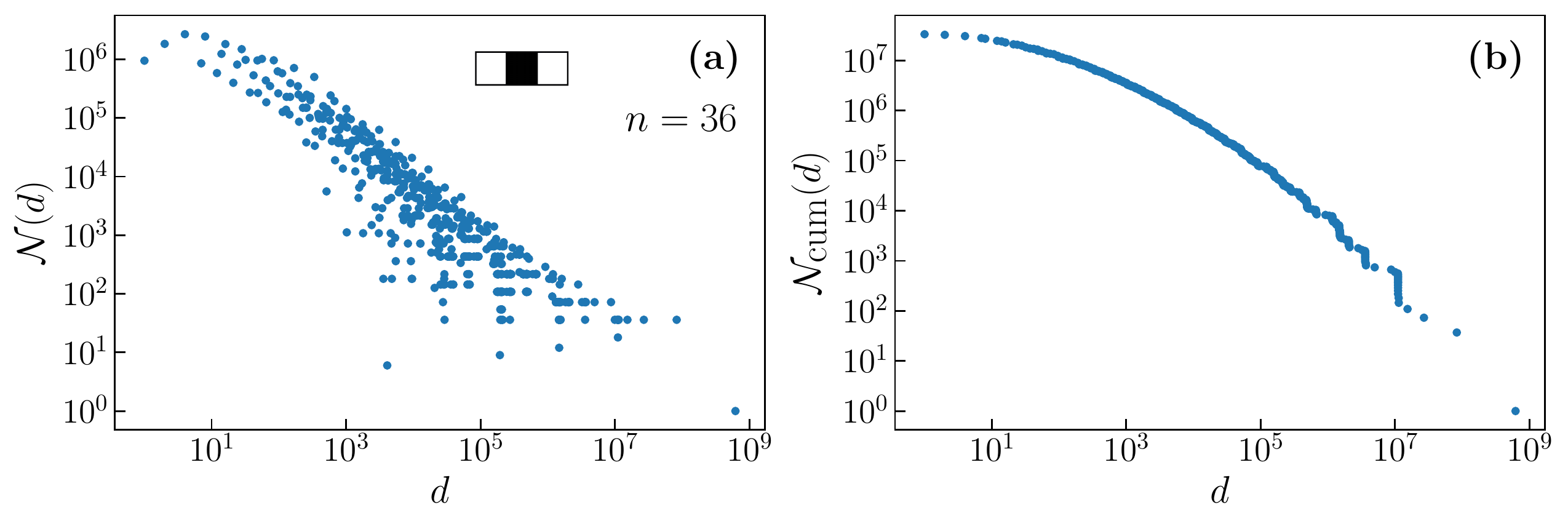}
\\
\includegraphics[width=0.68\textwidth]{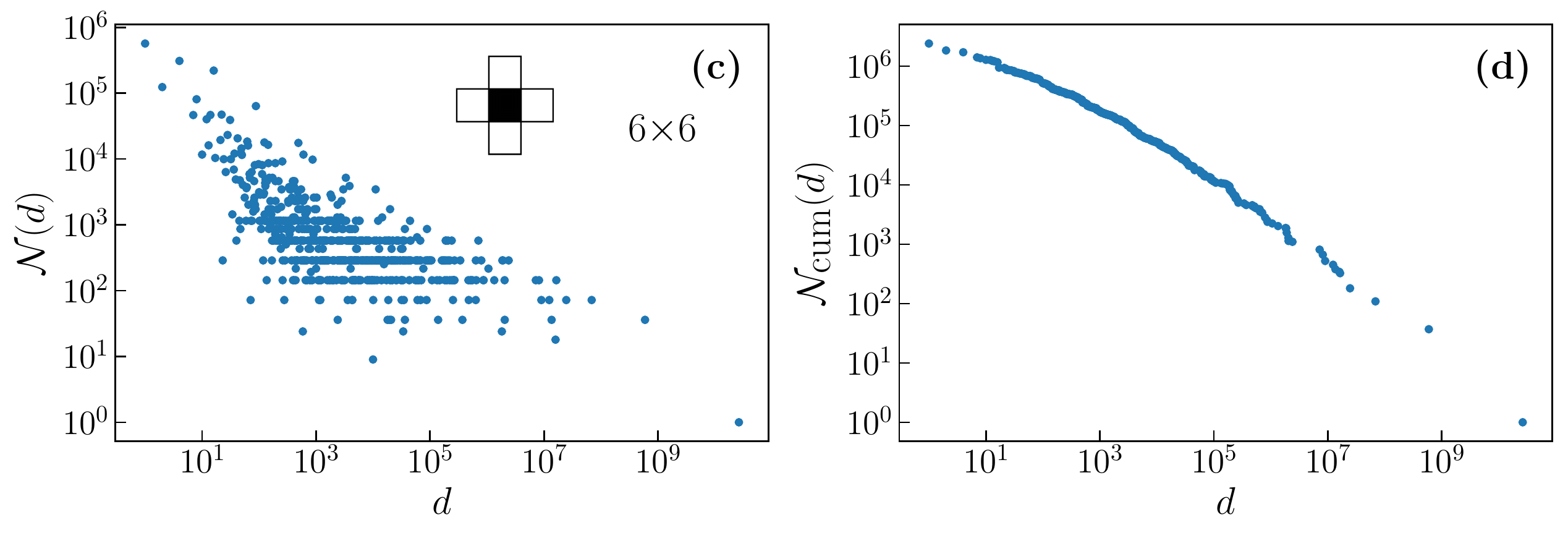}
\caption{Degeneracy distribution (a) and cumulative degeneracy distribution (b) for the filter 
$010$ on a ring,
and for its generalization on a torus, which is a 1 with four neighboring 0's [panels (c) and (d)].
}\label{distribution_examples}
\end{figure}

The filter works as follows: every instance of a specific pattern in the input (a short sequence of ones and zeroes) is marked by a one in the corresponding position in the output. All other positions are marked with zeroes. 
Multiple inputs correspond to the same output, creating a distribution of degeneracies of the outputs.
We illustrate the results from a simple example filter pattern in Figure \ref{distribution_examples} (a) and (b). We observe complex degeneracy distributions reminiscent of those observed in, for example, Ref. \cite{marsili2013sampling}.

The filter pattern may be arbitrary, but for illustrative purposes
we will consider in particular a family of filters consisting of a string of ones with zeroes at either end: $010$, $0110$, $01110$, etc.
The length of the filter, $w$, can be used as a crude control parameter to observe the effects on resolution and relevance (see below). For convenience, we use the notation $1_l$ to indicate a chain of $l$ ones. Thus the filter of length $w$ is $01_{w-2}0$.
In principle, for each of the $2^n$ possible inputs we can obtain, one by one, an output numerically. 
 In practice, we use a more efficient algorithm described in Ref.~\cite{baxter2020complex}. 
Other types of filter patterns on a ring may be analyzed using the same methods.

%%%=================================================================
%%%=================================================================
%%%=================================================================
%%%=================================================================

%%%%%%%%%%%%%%%
%%%%%%%%%%
%%%%%%%%%%%%%%
%%%%%%%%%%%
%%%%%%%%%%%%%%%
%%%%%%%%%%
%%%%%%%%%%%%%%

\subsubsection{Degeneracy distribution}

%\section{Output degeneracy distribution for complete input datasets}
%\label{s3pp} 

We obtained  the number of outputs ${\cal N}(d)$ for the full spectrum of degeneracies $d$ for a variety of filters.  The degeneracies $d_i$ , $i=1,...,D$, form a discrete spectrum of values where $d_D$ is the largest degeneracy, and $d_1=1$. 
A few examples of the degeneracy distributions and cumulative degeneracy distributions are shown in Figure~\ref{distribution_examples}.
Here ${\cal N}_\text{cum}(d_i) \equiv \sum_{j=i}^D {\cal N}(d_j)$. 
In particular, the total number of outputs is given by $M(n) = {\cal N}_\text{cum}(d_1)$. 
The cumulative degeneracy distribution is broad, but decays more rapidly than a power law.

The tail of the cumulative distribution has a notably complex structure
resembling a staircase, with steep jumps between steps.
The heights of these jumps are especially large in the region of high degeneracies. 
Similar structures may be observed in real systems, see for example Figure~3 of Ref. \cite{marsili2013sampling}.
As shown in Ref. \cite{baxter2020complex}, when the number of ones in the output is few, and some or all of them are separated by large gaps, such outputs have very similar but not exactly equal degeneracies for finite $n$. These closely located degeneracies lead to the staircase structure observed in the cumulative distribution.

%\subsubsection{Asymptotics of the degeneracy distribution}

Let us consider the evolution of the degeneracy distribution (and cumulative distribution) with input size $n$.
The largest degeneracy $d_D(n)$ corresponds to the output with all zeroes, 
and for large $n$, grows as $d_D(n) \cong z_d^n$, 
 where the value of $z_d$ depends on the specific filter. Naturally ${\cal N}( d_D , n) = 1$.
The number of outputs with degeneracy $1$ behaves as 
 ${\cal N}(1,n) \cong z_a^n$. 
Meanwhile the total number of outputs, $M(n)$ is asymptotically 
$M(n) \cong z_g^n$.
Together, these three key constants, $z_d$, $z_g$ and $z_a$, delimit the asymptotic behaviour of the degeneracy distribution \cite{baxter2020complex}. We list these numbers for a selection of short filter patterns on a ring in Table \ref{Table1}.

Rather surprisingly, one may obtain these asymptotic behaviours, and exact expressions for  the constants $z_g, z_d$ and $z_a$ through simple 
arguments. Each output consists of isolated ones separated by strings of zeroes of various lengths. By careful consideration of how valid outputs for a larger $n$ can be constructed by adding specific segments to shorter outputs, one may construct recursive relations for the key quantities $M(n)$, $d_D(n)$ and $N(1,n)$, whose asymptotics are given by  $z_g, z_d$ and $z_a$.
To demonstrate this, we focus on the particular family of filter patterns consisting of a chain of ones with a zero at each end. The shortest such pattern is $010$. Each member of this set may be indexed by the length of the filter, $w \geq 3$.
The filter pattern length $w$ determines the minimum number of zeroes, $w-2$, between each one. 
We give the derivation of $z_d, z_g$ and $z_a$ for any $w$ in Section \ref{MM_asymptotics} below.
%

 %%%%%%%%%%%%%%%%%%%%%%%%%%%%%%%%%%%%%%%%%%%%%%%%%%%%%%%%%%%%%%%%%%%%%%%%%%%%%%%%%%%%%%%%%%%%%%%%%%%%%%%%%%%%%%%%%%%%%%%%%%%%%%%%%%%%%%%%%%%%%%%%%%%%%%%%%%%%%%%%%%%%%%%%%%%%%%%%%%%%%%%%%%%%%%%%%%%%%%%%%%%%%%%%%%%%%%%%%%%%%%%%%%%%%%%%%%%%%%%%%%%%
 
 \subsubsection{Effect of filter length}

 %%%%%%%%%%%%%%%%%%%%%%%%%%%%%%%%%%%%%%%%%%%%%%%%%%%%%%%%%%%%%%%%%%%%%%%%%%%%%%%%%%%%%%%%%%%%%%%%%%%%%%%%%%%%%%%%%%%%%%%%%%%%%%%%%%%%%%%%%%%%%%%%%%%%%%%%%%%%%%%%%%%%%%%%%%%%%%%%%%%%%%%%%%%%%%%%%%%%%%%%%%%%%%%%%%%%%%%%%%%%%%%%%%%%%%%%%%%%%%%%%%%%%%%%%%%%%%%%%%%%%%%%%%%%%%%%%%%%%%%%%%%%%%%%%%%%%%%%%%%%%%%%%%%%%%%%%%%%%%%%%%%%%%%%%%%%%%%%%%%%%%%%%%%%%%%%%%%%%%%%%%%%%%%%%%%%%%%%%%%%%%%%%%%%%%%%%%%%%%%%%%%%%%%%%%%%%%%%%%%%%%%%%%%%%%%%%%%%%%%%%%%%%%%%%%%%%%%%%%%%%%%%%%%%%%%%%%%%%%%%%%%%%%%%%%%%%%%%%%%%%%%%%%%%%%%%%%%%%%%%%%%%%%%%%%%%%%%%%%%%%%%%%%%%%%%%%
\begin{table}%[H]
\caption{
Values of the numbers $z_g$, $z_d$, and $z_a$ for different filters. 
Note that we also included filter patterns consisting of all zeroes. 
For each filter we also give the relevance per node $H[d]/n$ (in nats) calculated from the degeneracy distribution and the resolution per node $H[y]/n$. For the sake of comparison, the standard entropy of the inputs of this size is $H/n=\ln 2 = 0.69315$. Finally we include the number of distinct degeneracies $D$ for each pattern. Inputs of size $n=36$ were used except for filters $00$ and $10$, for which $n=34$, and $000$ for which $n=35$. Values for $D$ for these three filters were extrapolated to $n=36$ for comparison with other results.
} \label{Table1}
\centering
%\tb newline within a bracket \ts new line of same size not bracketed, \tn new size
\begin{tabular}{p{0.1\textwidth} p{0.075\textwidth} p{0.025\textwidth} l l l l l c p{0.1\textwidth}}%c c c c c c p{0.1\textwidth}}
\toprule
& pattern && $z_g$ & $z_d$ & $z_a$ & $\!\!H[d]/n$ & $\!\!H[y]/n$ & $D$ &
\\ 
%\\ 
\midrule
& 0 (or 1) && 2 &  1 &  2 &   0 & 0.69315 &  1\tn
%
%%%%
& 00 && 1.75488 &  1.61803 &  1.61803 & 0.18261   &  0.48468  &  924(1)\ts
& 10  &&   1.61803 &   1.31951 & 1 & 0.13954 & 0.46986 & 513(1)\tn   %&  *(extrapolated from $n \leq 33$)\\
& \textbf{010}  && 1.61803 & 1.75488 & 1.46557 &  0.17248 & 0.35187 &  777\ts
& 000   &&1.61803 &   1.83929 & 1.49710 & 0.1453(1)  & 0.30105 & 554(2)\tn  %&  *(extrapolated from $n \leq 33$)\\
& \textbf{0110}  &  \multirow{2}{*}{$\left.\rule{0pt}{2.2ex}\right\}$} &   \multirow{2}{*}{1.46557} &   \multirow{2}{*}{1.86676} &   \multirow{2}{*}{1.22074} &   \multirow{2}{*}{0.11881} &   \multirow{2}{*}{0.22387} &   \multirow{2}{*}{698}\tb
& 0100\ts
& 0000 && 1.52895 &         1.92756 &      1.41963(2) &  0.08856 &	0.17673 & 311\tn
& 00100 &&      1.46557&    1.9417 &  1.38028 & 0.06434	&0.13371  &   291\ts
& \textbf{01110}   &   \multirow{3}{*}{$\left.\rule{0pt}{3.7ex}\right\}$} &    \multirow{3}{*}{1.38028} &    \multirow{3}{*}{1.93318} &    \multirow{3}{*}{1.16730} &   \multirow{3}{*}{0.06312} &   \multirow{3}{*}{0.13562} &    \multirow{3}{*}{255} \tb
& 01100\tb
& 01000\ts
& 01010 &&      1.44327  &  1.94789 &1.32472 & 0.06117	& 0.12584 &   301\ts
& 00000 &&     1.46557(2) & 1.96595 & 1.3652(2) & 0.05108 &	0.10052  &   190\tn
& 001100 &   \multirow{2}{*}{$\left.\rule{0pt}{2.2ex}\right\}$} &      \multirow{2}{*}{1.38028}  &     \multirow{2}{*}{1.96931} & \multirow{2}{*}{1.2499(2)}  & \multirow{2}{*}{0.03606} & \multirow{2}{*}{0.07899} & \multirow{2}{*}{197} \tb
& 001000\ts
& 010010 &&      1.37108(1)    &      1.97113 & 1.1938(5) & 0.03586	& 0.07692  &      218\ts
& \textbf{011110}  &   \multirow{6}{*}{$\left.\rule{0pt}{8.2ex}\right\}$} &   \multirow{6}{*}{ 1.32472} &   \multirow{6}{*}{1.96717} &    \multirow{6}{*}{1.13472}    &   \multirow{6}{*}{0.03448} &   \multirow{6}{*}{0.07939}  &   \multirow{6}{*}{123}\tb
& 011100\tb
& 011010\tb
& 011000\tb
& 010100\tb
& 010000\ts

& 000000 &&      1.4176(2) &     1.98358 & 1.32486 & 0.02968 &	0.05606 &         123\tn
& 0110110 &&   1.32472 & 1.98574 &  1.158(2)    & 0.02084 &	0.04353 & 129\ts
& \textbf{0111110} &\multirow{2}{*}{$\left.\rule{0pt}{2.2ex}\right\}$} &    \multirow{2}{*}{1.28520} &   \multirow{2}{*}{1.98386} &     \multirow{2}{*}{1.11278}  &  \multirow{2}{*}{0.02016} &  \multirow{2}{*}{0.04535} &  \multirow{2}{*}{64}\tb
& 0111010\tn% &&   & 1.28520(2) &    1.98386 &    1.110(5)    &   0.72561 & 1.63275   &  64\\[0.7ex]
& \textbf{01111110} &&   1.25542 &    1.99203 &     1.09698  &  0.01213 & 0.02546 &  36\tn
& \textbf{011111110} &&  1.23205 &    1.99605 &     1.08507  &  0.00727 & 	0.01411 &  25\tn
& \textbf{0111111110} && 1.21315 &    1.99803 &     1.07577  &  0.00427 & 0.00774 & 16\tn
\bottomrule
\end{tabular}
\end{table}
%%%%%%%%%%%%%%%%%%%%%%%%%%%%%%%%%%%%%%%%%%%%%%%%%%%%%%%%%%%%%%%%%%%%%%%%%%%%%%%%%%%%%%%%%%%%%%%%%%%%%%%%%%%%%%%%%%%%%%%%%%%%%%%%%%%%%%%%%%%%%%%%%%%%%%%%%%%%%%%%%%%%%%%%%%%%%%%%%%%%%%%%%%%%%%%%%%%%%%%%%%%%%%%%%%%%%%%%%%%%%%%%%%%%%%%%%%%%%%%%%%%%%%%%%%%%%%%%%%%%%%%%%%%%%%%%%%%%%%%%%%%%%%%%%%%%%%%%%%%%%%%%%%%%%%%%%%%%%%%%%%%%%%%%%%%%%%%%%%%%%%%%%%%%%%%%%%%%%%%%%%%%%%%%%%%%%%%%%%%%%%%%%%%%%%%%%%%%%%%%%%%%%%%%%%%%%%%%%%%%%%%%%%%%%%%%%%%%%%%%%%%%%%%%%%%%%%%%%%%%%%%%%%%%%%%%%%%%%%%%%%%%%%%%%%%%%%%%%%%%%%%%%%%%%%%%%%%%%%%%%%%%%%%%%%%%%%%%%%%%%%%%%%%%%%%%%

In analogy with complex systems, we can consider each filter pattern as sampling the hidden state of a complex system~\cite{baxter2020complex}. The length of these filters acts as a crude control parameter of our sampling. Intuitively, we expect shorter filters to be more informative. 
The resolution of a sampling process, defined as the entropy of a sample:
\begin{equation}
\label{resolution}
H[y] 
= -\frac{1}{N}\sum_{i=1}^N\log\left(\frac{d_i}{N} \right)
= -\sum_d \frac{d{\cal N}(d,n)}{N}\log\left(\frac{d}{N} \right)
\end{equation}
is a measure of the ability to distinguish, at the output, between different input states \cite{cubero2018minimally}. It takes its maximum value when there is a different output for each input.
However in this case all outputs are distinct, and so these filters are
not informative about the system being sampled.
As shown in Ref.~\cite{cubero2018minimally}, the informativeness of a sample is captured by a different entropy measure, the relevance, defined as
\begin{equation}
H[d] = -\sum_d \frac{d{\cal N}(d,n)}{N}\log\left(\frac{d{\cal N}(d,n)}{N} \right).\label{entropy}
\end{equation} 

Results for a variety of short filter patterns are given in  Table \ref{Table1}. The family of filters composed of a string of ones with a zero at each end, $010, 0110, 01110, $ etc., are indicated in boldface in the Table. % \ref{Table1}. 
As can be seen in the Table, the relevance is greater for shorter filters, but is actually zero for the shortest possible filters $0$ and $1$. 
The filter pattern $1$ trivially reproduces the input, while $0$ it's inverse, and all outputs have degeneracy one.
Within the family of filters $01_{w-2}0$, the relevance is maximised for $w=3$.

Filter patterns of length two begin to have nontrivial properties.
For the pattern $01$, the number of outputs with degeneracy $1$, $N(1,n)$, is either 0 (when $n$ is odd) or 2 (when $n$ is even), so $z_a = 1$. This is because the only outputs that have degeneracy one are periodic sequences of alternating $0$’s and $1$’s — there are two of these sequences $n$ is even, and none when $n$ odd. 
The maximum degeneracy $d_D(n)$ for this pattern grows by an integer factor of $4$ for an increment in $n$ of $5$. In fact it can be written explicitly,
\begin{equation}
d_D(n>11) = \left[\frac{3}{4^{4/5}}\right]^{-mod(n, -5)} 4^{n/5}
,
\end{equation}
where the coefficient of $4^{n/5}$ equals
$({3}/{4^{4/5}})^0=1$, $({3}/{4^{4/5}})^4=0.959164$, $({3}/{4^{4/5}})^3=0.969214$, $({3}/{4^{4/5}})^2=0.979369$, and $({3}/{4^{4/5}})^1=0.989631$ for $mod(n,5) = 0$, $1$, $2$, $3$, and $4$, respectively.
As a result, the number $z_d$, which gives the asymptotic behaviour of the maximum degeneracy $d_D$, is equal to $4^{1/5}$. 

As can be seen in Figure~\ref{graphs_distributions2}~(c), the degeneracy distribution of the filter $01$ does not have the characteristic shape, and the broad tailed cumulative distribution seen in other filters.
The filter pattern $00$ already produces more complexity, see Figure~\ref{graphs_distributions2} (a). The degeneracy distribution and the cumulative distribution already have the shape and complexity seen in longer filters~\cite{baxter2020complex}.
Curiously $ N(1,n) = d_D(n) + i^n+(-i)^n$ (where $i$ is the imaginary unit) where the last two terms give $0, 2, 0, -2, 0 , 2, 0, -2, 0,…$ for $n = 3, 4, 5, 6,…$. This means that $z_d = z_a \approx 1.618$.

The largest degeneracies behave as $\cong z_d^n$ for large $n$.
The number $z_d$ quickly approaches $2$ as the filter pattern length increases.  Since $N = 2^n$, this means that almost all outputs concentrate in a few outputs, and in the limit, in a single state, i.e. all outputs 
are the same 
and the filter patterns are not informative. For the shortest filter patterns, the value of $z_d$ falls rapidly, while the relevance increases, indicating a transition to informative sampling.
On the contrary, $z_g$, which gives the total number of outputs $M(n)$, increases with decreasing filter length, as shorter filters have more possible outputs. 
 Taken together, these results indicate that the maximally informative sampling for a given family of filters is the shortest pattern having length greater than $1$. This behavior is analogous to the transition observed in more complex problems (see for example \cite{cubero2018minimum}).

Note that one may also consider filters constructed as logical combinations of more than one pattern.
For example, there are 3 kinds of ‘OR’ filters of size $2 + 2$ (in fact, there are $(16-4)/2=6$ combinations of different filters, but some are equivalent in terms of degeneracies). 
All of these OR filters have trivial degeneracies:
$01$ OR $10$ detects when the next digit is different from the current one. Given the value of an input digit we can completely reconstruct that input, and, since the first input digit has 2 possible values, each output has degeneracy $2$. The only degeneracy in the spectrum is $2$ and its frequency is $2^{n-1}$, so $z_d=1$ and $z_g=2$. There are no outputs of degeneracy one, $N(1,n) = 0$.
$00$ OR $11$ detects when the next digit is the same as the current one. The same reasoning as for the filter $01$ OR $10$ applies here: we can reconstruct the input completely from the output if we know a single digit of the input.
Finally $11$ OR $10$ (which is the same as $11$ OR $01$, $00$ OR $10$ and $00$ OR $01$) is equivalent to the filter $1$ of length 1.

%%%%%%%%%%%%%%%%%%%%%%%%%%%%%%%%%%%%%%%%%%%%%%%%%%%%%%%%%%%%%%%%%%%%%%%%%%%%%%%%%%%%%%%%%%%%%%%%%%%%%%%%%%%%%%%%%%%%%%%%%%%%%%%%%%%%%%%%%%%%%%%%%%%%%%%%%%%%%%%%%%%%%%%%%%%%%%%%%%%%%%%%%%%%%%%%%%%%%%%%%%%%%%%%%%%%%%%%%%%%%%%%%%%%%%%%%%%%%%%%%%%%%%%%%%%%%%%%%%%%%%%%%%%%%%%%%%%%%%%%%%%%%%%%%%%%%%%%%%%%%%%%%%%%%%%%%%%%%%%%%%%%%%%%%%%%%%%%%%%%%%%%%%%%%%%%%%%%%%%%%%%%%%%%%%%%%%%%%%%%%%%%%%%%%%%%%%%%%%%%%%%%%%%%%%%%%%%%%%%%%%%%%%%%%%%%%%%%%%%%%%%%%%%%%%%%%%%%%%%%%%%%%%%%%%%%%%%%%%%%%%%%%%%%%%%%%%%%%%%%%%%%%%%%%%%%%%%%%%%%%%%%%%%%%%%%%%%%%%%%%%%%%%%%%%%%%

\subsection{Filtering on graphs}\label{graphs}

The process described in the previous Section may be generalised to an arbitrary graph as follows. 
The input consists of the binary status for each node in the graph.
We filter for a particular condition of the state of a node and of its immediate neighbors. If the state of the node and its neighbours matches the filter pattern, the output for that node is $1$, otherwise it is $0$.
We consider two examples: Firstly, we set the output to $1$ if the selected node has state $1$ and all of its neighbours have state $0$ (we refer to this filter as the strong rule, or SR). This filter applied on a ring is equivalent to the pattern $010$ discussed in the previous Section. 
Secondly, we apply a less selective filter, outputting $1$ if a node is in state $1$ and {\em any} of its neighbours has state $0$ (we call this filter the weak rule, or WR).
We illustrate the application of these two filter patterns to a small graph in Figure~\ref{graph_fig}.

These filters were applied to several families of degree-regular graphs. These were chosen to have a variety of structures and to vary in the degree of randomness in their construction, while being of comparable size and degree.
We considered the following families of graphs:
Small world graphs. These graphs created by placing all nodes in a ring, and adding shortcuts between nodes to reach the desired degree. The locations of shortcuts were either random -- we use the code SW(q) for these graphs, where $q$ is the graph degree -- or in a deterministic way -- SWB(q);
Random regular graphs (RRG);
Tori, which are two dimensional square lattices with cyclic boundary conditions;
Cages. These are graphs defined by two numbers, the degree $q$ and the shortest cycle length $g$. A (q,g)-cage is the graph fulfilling these 
properties while having the smallest possible numbers $n$ of nodes \cite{meringer1999fast}. 
For each family of graphs we considered different sizes, up to at least $n=30$, and where possible, degrees, from $q=2$ up to $q=5$.
Finally we investigated the second and third
Apollonian networks (Apollonian 2 and 3), which are the only graphs here that are not degree regular.

 %%%=================================================================
 %%%=================================================================
 \begin{figure}[H]
 \centering
 \includegraphics[width=0.7\textwidth]{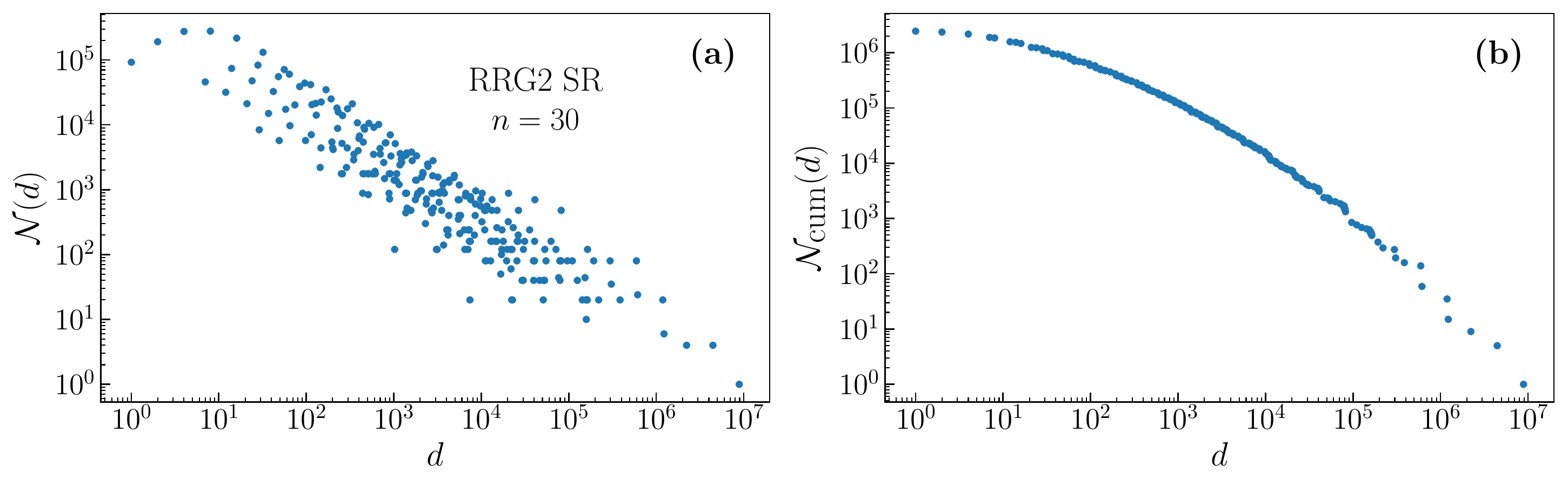}
 \includegraphics[width=0.7\textwidth]{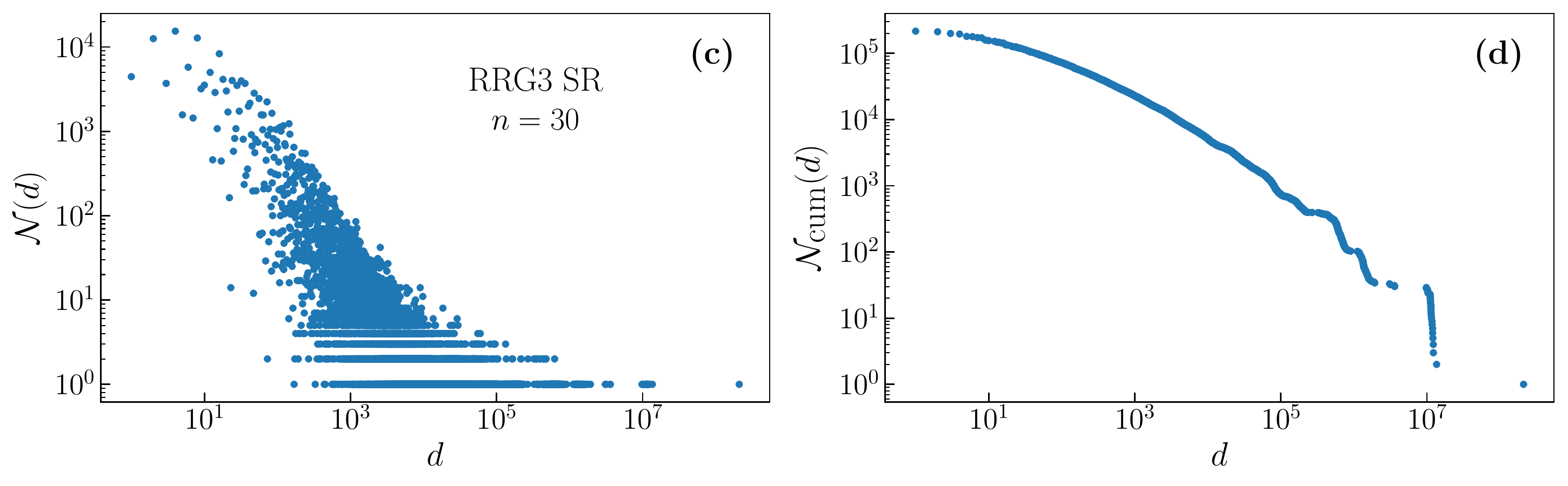}
 \includegraphics[width=0.7\textwidth]{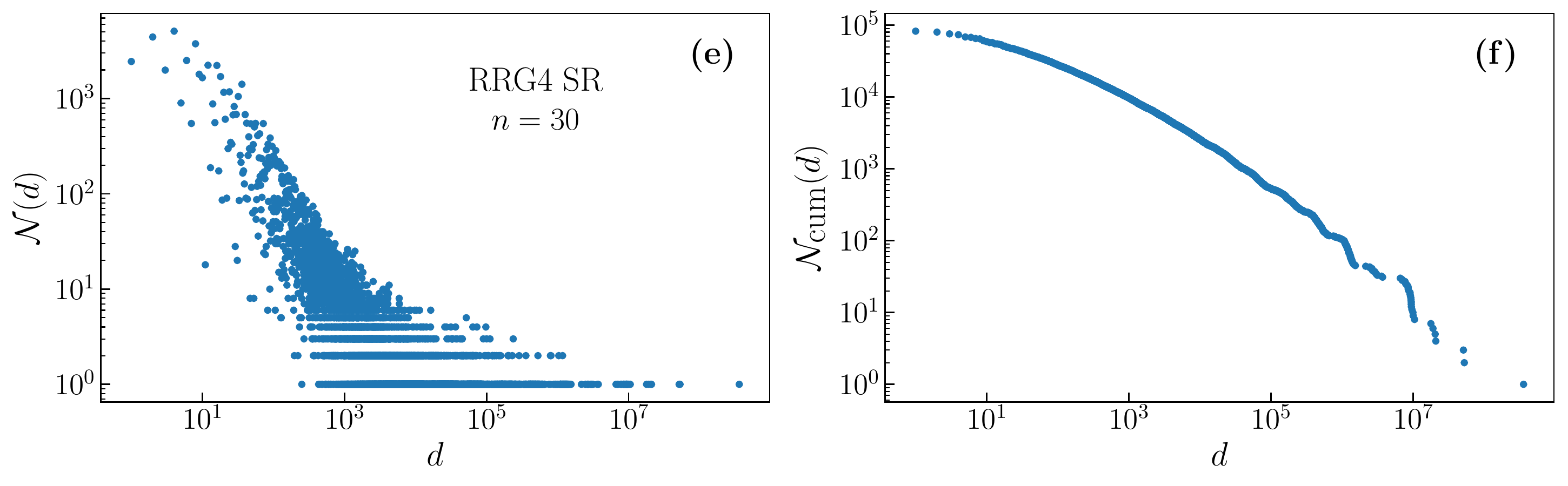}
 \caption{
 Degeneracy distributions (left) and cumulative degeneracy distributions (right) for outputs of the SR filter on random regular graphs of degree 2 (a,b) 3 (c,d) and 4 (e,f).
 %and 5 (g,h)
 }\label{graphs_distributions}
 \end{figure}   
 %%%=================================================================
 %%%=================================================================

 %%%=================================================================
 %%%=================================================================
 \begin{figure}[H]
 \centering
 \includegraphics[width=0.7\textwidth]{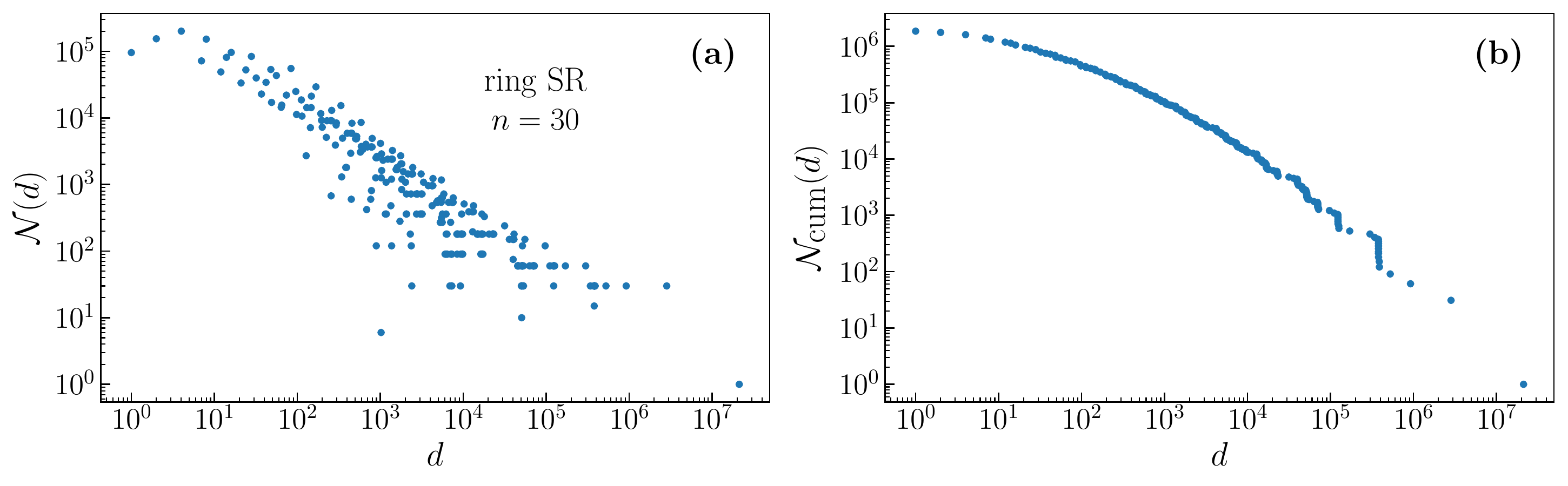}
 \includegraphics[width=0.7\textwidth]{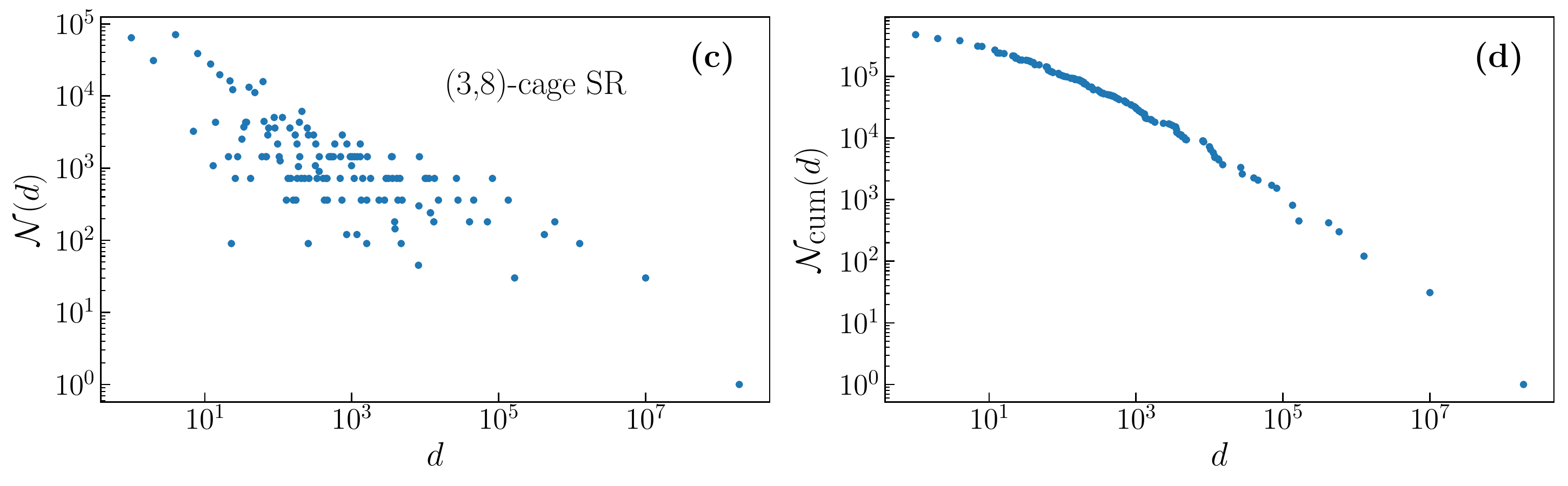}
 \includegraphics[width=0.7\textwidth]{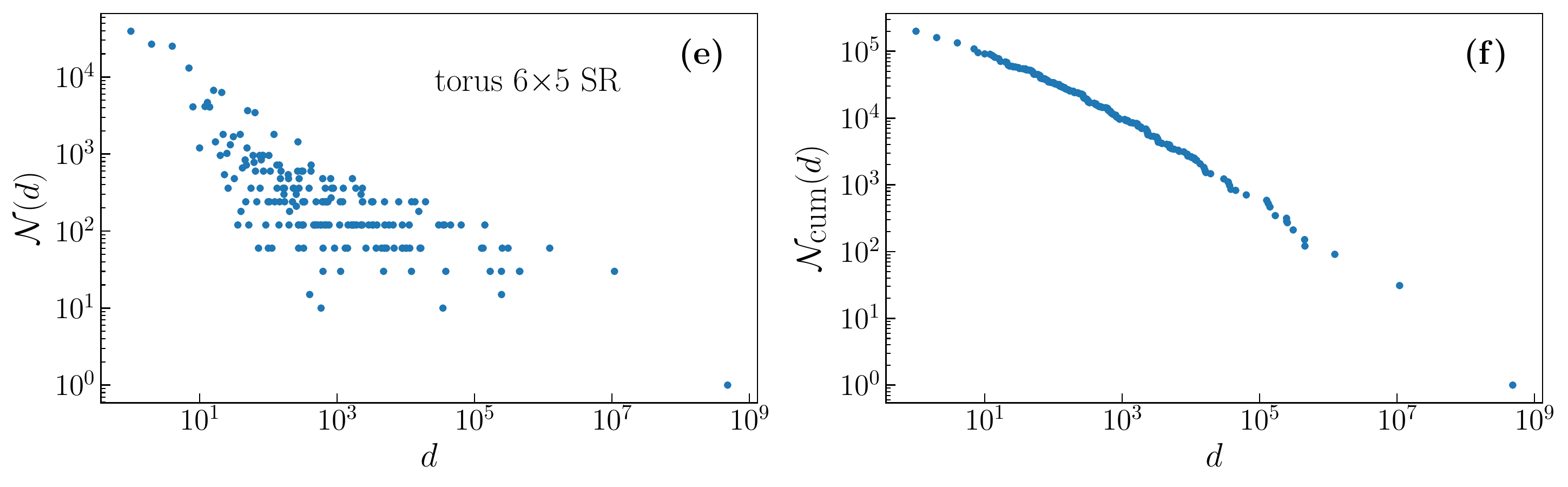}
 \caption{
  Degeneracy distributions and cumulative degeneracy distributions for outputs of the SR filter on selected deterministic graphs of degree 2 (a,b) 3 (c,d) and 4 (e,f).
 }\label{graphs_distributions1}
 \end{figure}   
 %%%=================================================================
 %%%=================================================================

We give some examples of the resulting degeneracy distributions and cumulative degeneracy distributions, for the SR filter, in Figures~\ref{graphs_distributions}, for random graphs, and \ref{graphs_distributions1}, for deterministically generated graphs.
Note that the distributions for random graphs correspond to a single realization of the graph. We see that there is a dramatic difference in the distribution for random graphs between degree two and degree three. The degree two random regular graph necessarily consists of one or several closed rings, and the distribution is little different than that shown in Figure~\ref{distribution_examples} (a). For degree three, there is a great deal of randomness in the formation of the graph, and this is reflected in the degeneracy distribution, which becomes much more dense, having a fine structure not observed in deterministic graphs. For higher degrees, the distribution becomes less broad, and as we will discuss below, this corresponds to a reducing relevance with increasing degree.

We have not included examples of the distributions for the "small world" graphs. The deterministic small world graphs, SWB(q), produce distributions almost indistinguishable from those for other deterministic graphs of the same degree, while the random small world graphs, SW(q), generate degeneracy distributions very similar to those found for random regular graphs.
For completeness, we give the degeneracy distributions and cumulative distributions for the same graphs using the WR filter in Figures~\ref{graphs_distributions4} and \ref{graphs_distributions5} in Appendix \ref{APP_WRresults}.

 %%%=================================================================
 %%%=================================================================
 \begin{figure}[H]
 \centering
 \includegraphics[width=0.7\textwidth]{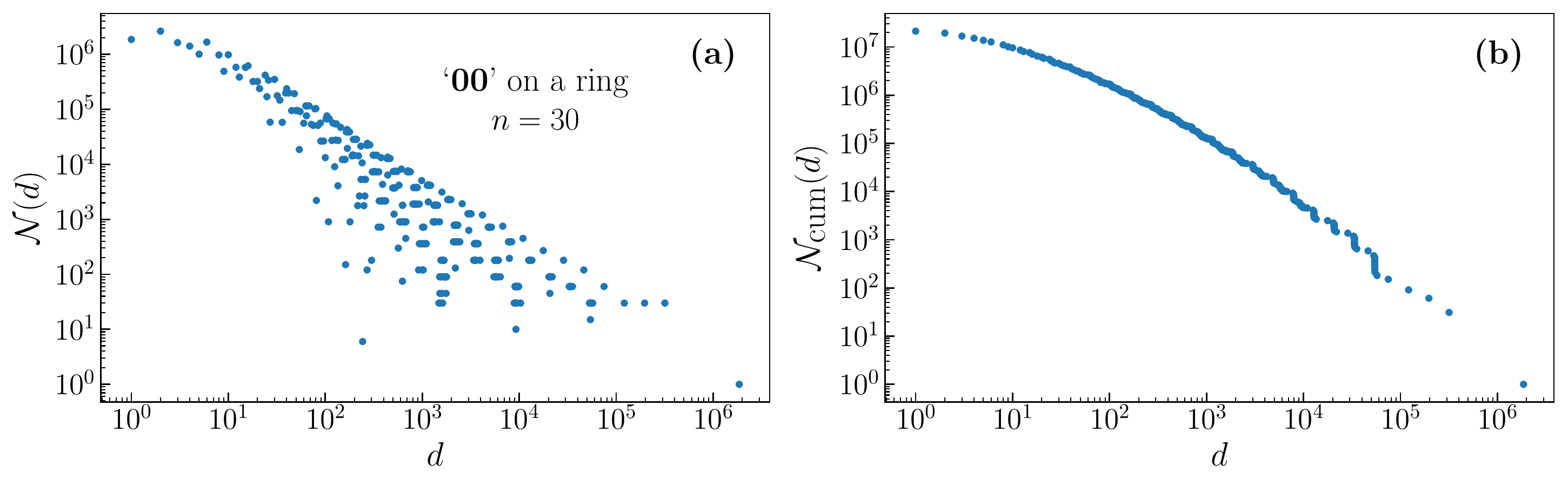}
 \includegraphics[width=0.7\textwidth]{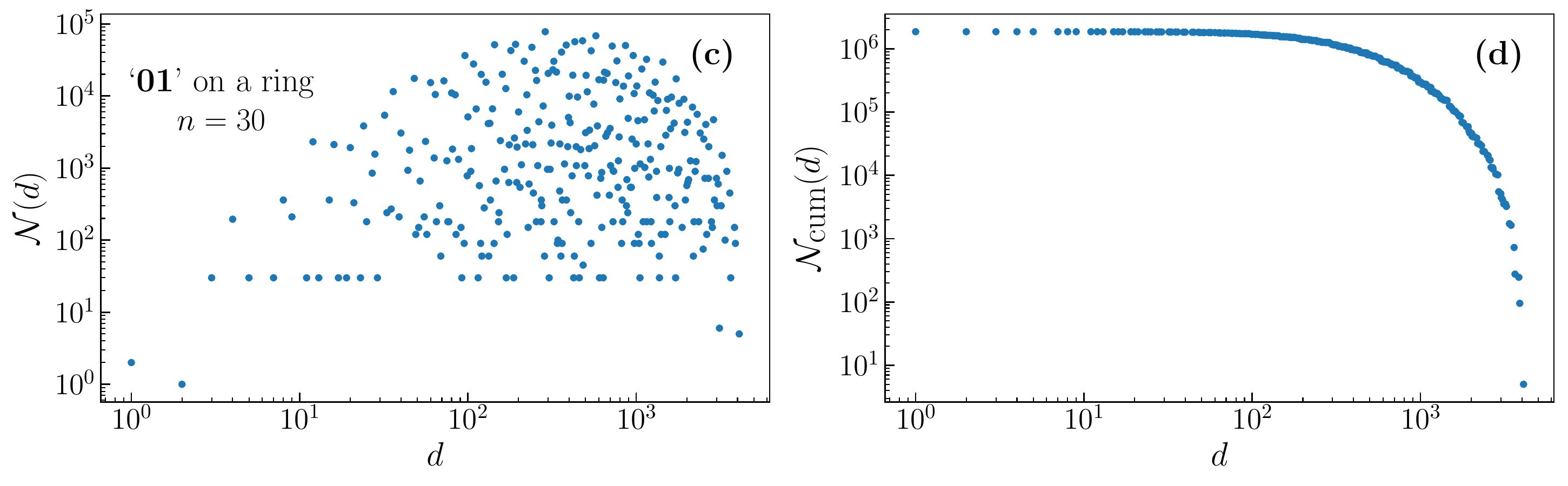}
 \includegraphics[width=0.7\textwidth]{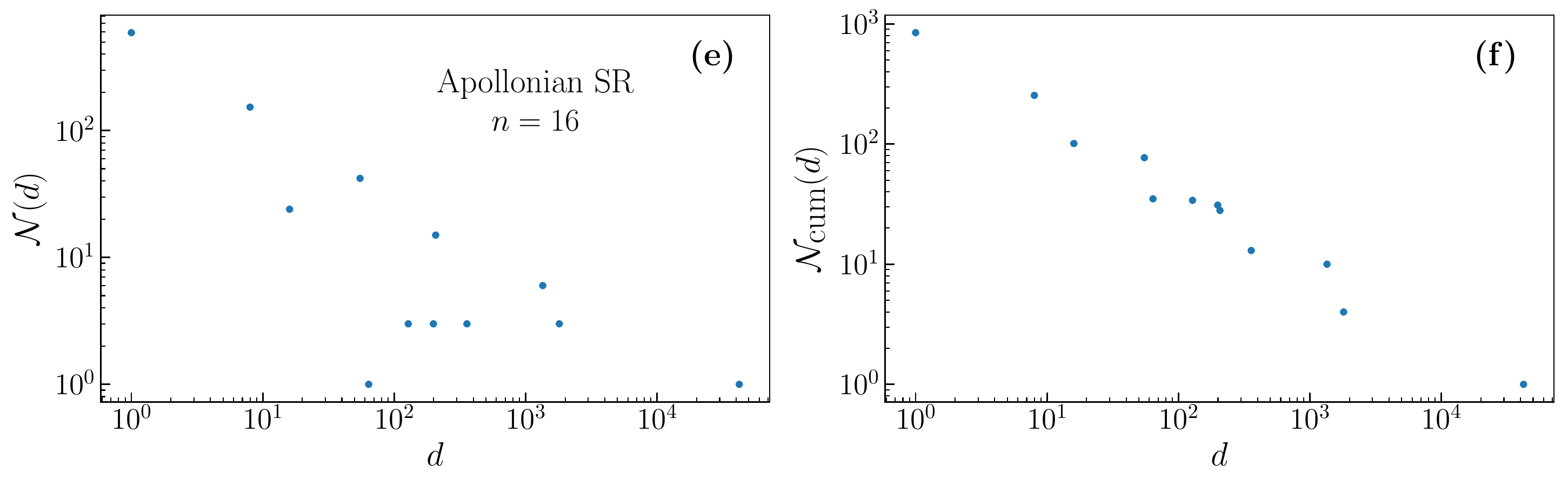}
 \includegraphics[width=0.7\textwidth]{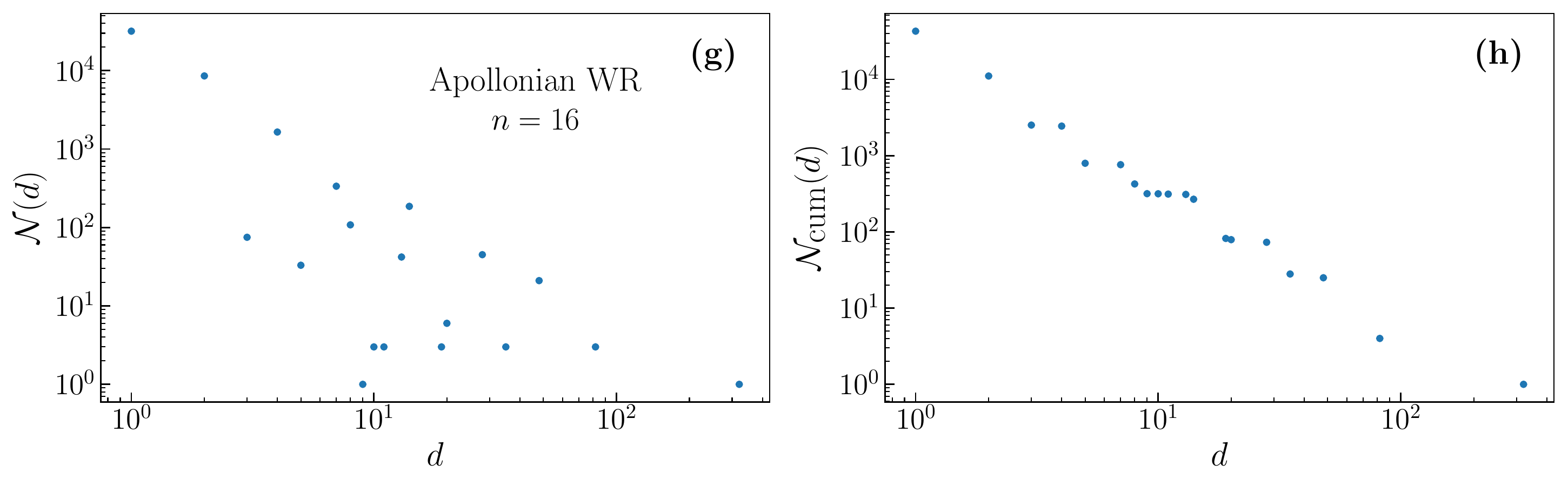}
 \caption{
  Degeneracy distributions and cumulative degeneracy distributions for outputs of the SR filter on selected deterministic graphs of degree 2 (a,b) 3 (c,d) and 4 (e,f).
 }\label{graphs_distributions2}
 \end{figure}   
 %%%=================================================================
 %%%=================================================================

We plot some examples of some less typical degeneracy distributions in Figure~\ref{graphs_distributions2}. These are the $00$ and $10$ filters applied on a ring, and the SR filter applied to Apollonian networks (which are not degree regular).

%%%=================================================================
%%%=================================================================
\begin{figure}[H]
\centering
\includegraphics[width=0.8\textwidth]{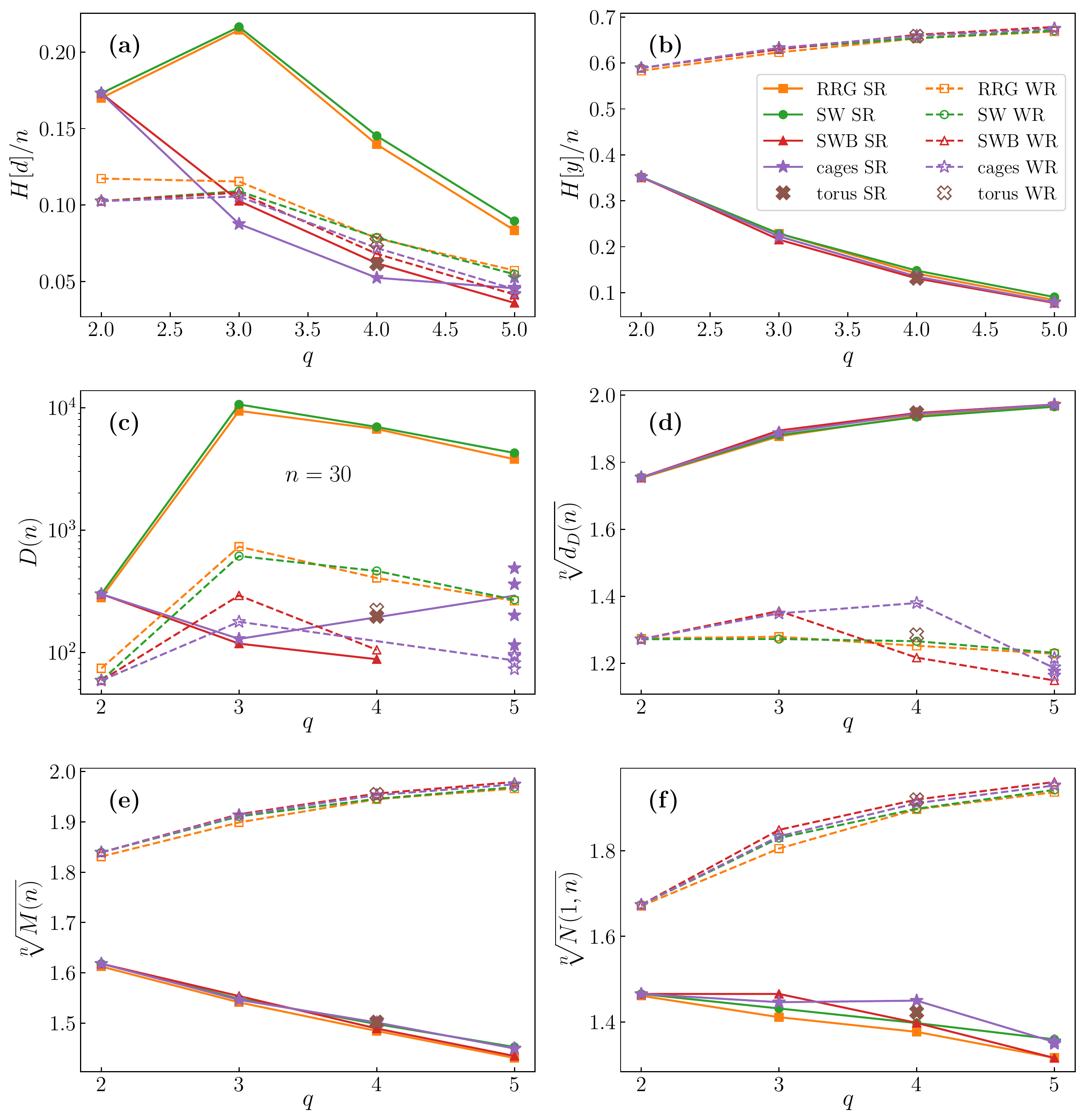}
\caption{Dependence of key observables related to the degeneracy distribution on graph degree $q$.
(a) The relevance entropy $H[d]$ scaled by system size $n$.
(b) Resolution $H[y]$.
(c) Total number of degeneracies $D(n)$ for $n=30$.
(d) The $n$\textsuperscript{th} root of the largest degeneracy $d_D(n)$, which tends to $z_d$.
(e) The $n$\textsuperscript{th} root of the number of outputs $M(n)$, tending to $z_g$.
(f) The $n$\textsuperscript{th} root of the number of outputs of degeneracy onem $N(1,n)$, tending to $z_a$.
}\label{numbers_vs_q}
\end{figure}   
%%%=================================================================
%%%=================================================================

In Figure~\ref{numbers_vs_q} we represent various quantities of interest as a function of graph degree, for the different graph families studied. We see that there is a clear separation in results between the two filters. 

The weak filter (WR) detects when a node has state $1$ while having at least one immediate neighbor with state $0$. This neighbour condition is more easily satisfied the larger the number of neighbours $q$. Thus for large $q$, the number of possible outputs $M(n)$ for the WR filter approaches the number of possible inputs, $2^n$. We see in panel (e) that indeed the $n${\textsuperscript{th}} root of $M(n)$, which tends to $z_g$ for large $n$, approaches $2$ for large $q$. By the same token, most outputs have a degeneracy of one, so the number of outputs of degeneracy one, $N(1,n)$ also approaches $2^n$ ($z_a$ approaching $2$) for large $q$ [panel (f)], with while the largest degeneracy $d_D(n)$ (whose asymptotic behaviour is given by $z_d$) grows only slowly with $n$, [panel (d)].
The resolution $H[y]$ measures how well the filter distinguishes different inputs, and as we see in panel (b) of Figure~\ref{numbers_vs_q}, and in agreement with the above observations, the resolution for the weak filter is high. The maximum possible value of $H[y]$ is $n \ln 2$, corresponding to a value of $H[y]/n = 0.693...$ in the figure. We see that the resolution is already close to this value at $q=5$.

The correct measure of how informative a sample of the observable variables of a complex system is about the underlying system is the relevance \cite{cubero2018minimally}, $H[d]$. Such sampling is represented in our problem as the filtering process, and the interactions of the system by the graph structure. The importance of the relevance is confirmed by our results, as shown in Figure~\ref{numbers_vs_q} (a). A higher relevance is measured in graphs having some randomness in their structure, while deterministic and regular graphs have lower relevance. This is particularly true for the strong filter SR, which produces a significantly higher relevance for random regular graphs (RRG) and rings with random shortcuts (SW), compared with rings with deterministic shortcuts (SWB) and cages. The effect for the WR filter is much less pronounced.

The highest relevance occurs at degree $q=3$. The explanation for this is clear. As shown above, and in \cite{baxter2020complex}, smaller filters generally produce higher relevance, as there are more outputs than for larger filters, except in the extreme limit of perfect reproduction of the input (maximum resolution). Thus we would expect lower values of $q$, which correspond to smaller SR filters, to have higher relevance. Meanwhile, and opposing this trend, graphs of degree $q=2$ are necessarily either rings or sets of rings, which thus have a (nearly) deterministic structure and suffer a penalty in relevance. Notice also the similarity of the degeneracy distributions for $q=2$ in Figures~\ref{graphs_distributions} and \ref{graphs_distributions4}. As can be seen in the figure, the reduction in relevance in moving from $q=3$ to $q=2$ due to this regularity outweighs the expected increase due to the filter being smaller. To put it another way, the maximum relevance occurs at the smallest value of $q$ for which the graph is non deterministic. This echoes our finding for filters on rings, for which the maximum relevance is found for the shortest filter which doesn't trivially reproduce the input \cite{baxter2020complex}.
We show the degeneracy ditribution for $q=3$ for a deterministic graph in Figure~\ref{graphs_distributions1}, and for a random graph in Figure~\ref{graphs_distributions}.

For the SR filter, in contrast to the weak filter, there is significant degeneracy of the outputs. The number of outputs is significantly less than the number of inputs, as is the number of outputs with degeneracy one. Similarly, the resolution is small for the SR filter, for all graph families, and decreases with $q$. The largest degeneracy, $d_D$, on the other hand, does become very large. In the limit of large $q$, a large fraction of possible outputs give the same single output (all zeroes).
In Figure~\ref{numbers_vs_q} (c), the behaviour of the number of degeneracies, $D(n)$ noticeably mirrors that of the relevance, $H[d]$.
Note that data points for random graphs are averaged over several realisations of the graph.

In Table \ref{Table2} we list the key degeneracy distribution statistics for the SR filter, for all families of graphs studied. 
Corresponding results for the WR filter may be found in Table \ref{Table3}.
In addition to representing the data highlighted in Figure~\ref{numbers_vs_q} in the quantitative form, these tables demonstrate the size effects with exponentially rapid convergence to the infinite $n$ limit. 
In this work, we are mainly interested in regular graphs (graphs where nodes have a uniform degree), because we can better isolate the effects of varying the graph's degree.
Nevertheless, for the sake of completeness, we also present results for a few examples of non-regular graphs, namely Apollonian networks.
In Tables \ref{Table2} and \ref{Table3}, each group of rows delimited by horizontal lines represents a different class of graphs.
The four classes at the top of the tables, namely Apollonian networks, cage graphs, square lattices with periodic boundary conditions (torus), rings with deterministic shortcuts, are deterministic graphs, while the two remaining classes represent random models, namely random regular graphs, and rings with random shortcuts.
The numbers presented for the random models result from averaging over 10 realizations sampled uniformly at random.

We include results for graphs of several sizes for each type of graph. This allows one to see the convergence of values with increasing $n$. 
Within the set of consecutive rows of each class, the graphs are ordered by ascending degree, then by ascending number of nodes.
The exception to this organization is the two first rows, which are for the non-regular Apollonian networks.
All of these numbers, as well as the number of degeneracies $D$, for $n=30$ are plotted in Figure~\ref{numbers_vs_q}.

%%%%%%%%%%%%%%%%%%%%%%%%%%%%%%%%%%%%%%%%%%%
%%%%%%%%%%%%%%%%%%%%%%%%%%%%%%%%%%%%%%%%%%%
%%%%%%%%%%%%%%%%%%%%%%%%%%%%%%%%%%%%%%%%%%%
\begin{table}[H]
\caption{
Important values for the degeneracy distribution resulting from applying the strong rule (SR) filter to various graphs.
The numbers $\sqrt[{\scriptstyle n}]{M(n)}$,  $\sqrt[{\scriptstyle n}]{d_D(n)}$ and  $ \sqrt[{\scriptstyle n}]{N(1,n)}$ approximate $z_g$,  $z_d$ and $z_a$ respectively. We also give the relevance per node $H[d]/n$ and the resolution per node $H[y]/n$.
Numbers for RRG($q$) and SW($q$) were obtained by averaging over 10 random realizations.
}
\label{Table2}
\centering
%\tb newline within a bracket \ts new line of same size not bracketed, \tn new size
\begin{tabular}{l c l l c c c p{0.1\textwidth}}%c c c c c c p{0.1\textwidth}}
\toprule    
%graph         nodes(n)     H[d] (nats)             H[y] (nats) & # degeneracies[D(n)] & max degen[d_D(n)]    # outputs[M(n)]    # outputs of d=1[N(1,n)]
%
graph         & $n$  &  $\sqrt[{\scriptstyle n}]{M(n)}$   & $\sqrt[{\scriptstyle n}]{d_D(n)}$    & $\sqrt[{\scriptstyle n}]{N(1,n)}$ &   $H[d]/n$     &   $H[y]/n$  \\
\midrule
Apollonian 2  & 7&  	1.47236	 & 	1.94420	 & 	1.40854	 & 	0.08504	 & 	0.12919	   \\
Apollonian 3  & 16 &  	1.52380	 & 	1.94596	 & 	1.49013	 & 	0.08148	 & 	0.13005	   \\
\midrule
(3,5)-cage & 10 &  	1.54199	 & 	1.88916	 & 	1.42694	 & 	0.10463	 & 	0.22185	   \\
(3,6)-cage &  14 &  	1.54904	 & 	1.88549	 & 	1.46952	 & 	0.09741	 & 	0.22302	   \\
(3,7)-cage &  24 &  	1.54516	 & 	1.88688	 & 	1.42191	 & 	0.12412	 & 	0.22268	   \\
(3,8)-cage & 30 &  	1.54618	 & 	1.88722	 & 	1.44630	 & 	0.08763	 & 	0.22254	   \\
(4,5)-cage & 19 &  	1.48991	 & 	1.94458	 & 	1.37494	 & 	0.08094	 & 	0.13458	   \\
(4,6)-cage & 26 &  	1.50129	 & 	1.94386	 & 	1.44997	 & 	0.05243	 & 	0.13497	   \\
(5,5)-cage 1 & 30 &  	1.44928	 & 	1.97192	 & 	1.34932	 & 	0.04164	 & 	0.07890	   \\
(5,5)-cage 2 & 30 &  	1.44984	 & 	1.97191	 & 	1.35558	 & 	0.04602	 & 	0.07891	   \\
(5,5)-cage 3 & 30 &  	1.44954	 & 	1.97192	 & 	1.35543	 & 	0.04201	 & 	0.07890	   \\
(5,5)-cage 4 & 30 &  	1.44964	 & 	1.97191	 & 	1.35280	 & 	0.05264	 & 	0.07891	   \\
\midrule
torus $3{\times}3$ & 9 &  		1.47967	 & 	1.94480	 & 	1.42350	 & 	0.07165	 & 	0.13112	   \\
torus $4{\times}4$ &  16 &  	1.51160	 & 	1.94843	 & 	1.46895	 & 	0.06205	 & 	0.13043	   \\
torus $5{\times}5$ &  25 &   	1.50066	 & 	1.94752	 & 	1.41779	 & 	0.05857	 & 	0.13132	   \\
torus $6{\times}5$ &  30 &  	1.50206	 & 	1.94754	 & 	1.42286	 & 	0.06159	 & 	0.13131	   \\
torus $10{\times}3$ &  30 &  	1.48922	 & 	1.94678	 & 	1.39796	 & 	0.05933	 & 	0.13100	   \\
torus $8{\times}4$ &  32 &	  	1.50701	 & 	1.94785	 & 	1.44980	 & 	0.06251	 & 	0.13096	   \\
torus $6{\times}6$ &  36 &  	1.50405	 & 	1.94756	 & 	1.44490	 & 	0.05890	 & 	0.13130	   \\
\midrule
SWB(3) & 10 &   	1.55564	 & 	1.89336	 & 	1.48457	 & 	0.10987	 & 	0.21539	   \\
SWB(3) & 20 &   	1.55376	 & 	1.89450	 & 	1.46394	 & 	0.09850	 & 	0.21540	   \\
SWB(3) & 30 &   	1.55377	 & 	1.89450	 & 	1.46573	 & 	0.10256	 & 	0.21541	   \\
SWB(4) & 12 &   	1.48818	 & 	1.94653	 & 	1.40063	 & 	0.07107	 & 	0.13103	   \\
SWB(4) & 21 &   	1.48924	 & 	1.94678	 & 	1.39802	 & 	0.06401	 & 	0.13100	   \\
SWB(4) & 30 &   	1.48922	 & 	1.94678	 & 	1.39797	 & 	0.06211	 & 	0.13100	   \\
SWB(5) & 12 &   	1.43618	 & 	1.97359	 & 	1.32007	 & 	0.04433	 & 	0.07602	   \\
SWB(5) & 20 &   	1.43469	 & 	1.97223	 & 	1.31634	 & 	0.03927	 & 	0.07765	   \\
SWB(5) & 32 &   	1.43463	 & 	1.97225	 & 	1.31607	 & 	0.03597	 & 	0.07765	   \\
\midrule
RRG(2) & 10 &   	1.55934	 & 	1.77122	 & 	1.41900	 & 	0.15869	 & 	0.32044	   \\
RRG(2) & 20 &   	1.60061	 & 	1.76297	 & 	1.45744	 & 	0.16195	 & 	0.33977	   \\
RRG(2) & 30 &   	1.61251	 & 	1.75289	 & 	1.46125	 & 	0.16997	 & 	0.35053	   \\
RRG(3) & 10 &   	1.49614	 & 	1.87903	 & 	1.30837	 & 	0.17514	 & 	0.21708	   \\
RRG(3) & 20 &   	1.52503	 & 	1.87847	 & 	1.37793	 & 	0.20373	 & 	0.22357	   \\
RRG(3) & 30 &   	1.54129	 & 	1.87706	 & 	1.41134	 & 	0.21442	 & 	0.22868	   \\
RRG(4) & 10 &   	1.44023	 & 	1.93770	 & 	1.30201	 & 	0.11648	 & 	0.13463	   \\
RRG(4) & 20 &   	1.48205	 & 	1.93399	 & 	1.36490	 & 	0.14077	 & 	0.14659	   \\
RRG(4) & 30 &   	1.48439	 & 	1.93797	 & 	1.37705	 & 	0.13959	 & 	0.14166	   \\
RRG(5) & 10 &   	1.41641	 & 	1.95111	 & 	1.27098	 & 	0.10038	 & 	0.11042	   \\
RRG(5) & 20 &   	1.42488	 & 	1.96513	 & 	1.30706	 & 	0.08722	 & 	0.08896	   \\
RRG(5) & 30 &   	1.43068	 & 	1.96825	 & 	1.31653	 & 	0.08344	 & 	0.08393	   \\
\midrule
SW(3) & 10 &   	1.55356	 & 	1.87107	 & 	1.43216	 & 	0.16610	 & 	0.23788	   \\
SW(3) & 20 &   	1.55998	 & 	1.86334	 & 	1.43951	 & 	0.22050	 & 	0.24702	   \\
SW(3) & 30 &   	1.54842	 & 	1.88077	 & 	1.43167	 & 	0.21643	 & 	0.22839	   \\
SW(4) & 10 &   	1.47637	 & 	1.91449	 & 	1.33514	 & 	0.14321	 & 	0.16987	   \\
SW(4) & 20 &   	1.49157	 & 	1.93385	 & 	1.38141	 & 	0.14235	 & 	0.14885	   \\
SW(4) & 30 &   	1.49811	 & 	1.93505	 & 	1.39764	 & 	0.14519	 & 	0.14804	   \\
SW(5) & 10 &   	1.43017	 & 	1.95045	 & 	1.29919	 & 	0.09802	 & 	0.11240	   \\
SW(5) & 20 &   	1.44575	 & 	1.95995	 & 	1.33955	 & 	0.09722	 & 	0.09962	   \\
SW(5) & 30 &  	1.45251	 & 	1.96562	 & 	1.35962	 & 	0.08951	 & 	0.09047	   \\
\bottomrule
\end{tabular}
\end{table}
%%%%%%%%%%%%%%%%%%%%%%%%%%%%%%%%%%%%%%%%%%%
%%%%%%%%%%%%%%%%%%%%%%%%%%%%%%%%%%%%%%%%%%%
%%%%%%%%%%%%%%%%%%%%%%%%%%%%%%%%%%%%%%%%%%%

%%%%%%%%%

For fully connected graphs, both the strong and the weak rules produce trivial output and degeneracy distributions.
Using the strong rule, for an output node $y_i$ to be 1, we must have $x_i=1$ and all other inputs $x_{j\neq i}=0$.
So, when there is a 1 in the output string, we have $y_i=x_i$. There are $n$ of these outputs, and their degeneracy is 1.
Since, there can be no more than a single $1$ in the output string, the only other possible output is a string of $n$ zeros, which has degeneracy $2^n-n$.
In this case there are only two degeneracies in the degree distribution $d_1=1$ and $d_2=2^n-n$, and their frequencies are $N(d_1,n)=n$, and $N(d_2,n)=1$, respectively.

On a fully connected graph under the weak rule, for an output node $y_i$ to be 1, it is enough to have $x_i=1$ and just one other input $x_{j\neq i}=0$.
Therefore, when one or more of the inputs $x_i$ is 0 the output is equal to the input.
The only situation in which the output does not match the input is for an input string of all 1’s, in which case the output is a strings of 0’s.
The weak rule also produces only two degeneracies $d_1=1$ and $d_2=2$, with frequencies $N(d_1,n)=2^n-2$, and $N(d_2,n)=1$.

It is worth noticing the relation with the class of cage graphs, which we have studied here: $(q,3)$-cage graphs are fully connected graphs with $q+1$ nodes, while $(q,4)$-cages are bipartite graphs with two fully connected layers of $q$ nodes each. Bipartite graphs with two fully connected layers of the same size also result in trivial degeneracy distributions in both the strong and weak rules.
With the strong rule applied to such a bipartite graph, for an output $y_i$ to be 1 we must have all inputs in the opposite layer to be $x_i=0$.
Conversely, when one input of one of the layers is $x_i=1$ all the outputs of the other layer are 0.
So, when all the input digits of one of the layers are all equal to 0 the outputs equal the inputs, $y_i=x_i$, and when there are 1’s in both layers of the input, the output is all 0’s.
In this case the degeneracy distribution also contains just two degeneracies, $d_1=1$ and $d_2=2^n-2^{n/2+1}$, with frequencies $N(d_1,n)=2^{n/2+1}$ and $N(d_2,n)=1$, respectively (notice there are $n/2$ nodes in each layer).
With the weak rule applied to symmetrical fully connected bipartite graphs, for an output $y_i$ to be 1 it is enough to have just one $x_i=0$ in the opposite layer.
Therefore, all inputs with at least a 0 in each layer produce an output $y_i=x_i$.
All inputs with at least one 0 in layer $\alpha$ and only 1’s in layer $\beta$ produce an output consisting of all 0’s in layer $\alpha$ and all 1’s in layer $\beta$.
Finally, if the input contains no 0’s in either layer, the output is $y_i=0$ for all $i$.
Therefore, we have $d_1=1$, $d_2=2$, and $d_3=2^{n/2}-1$, with frequencies $N(d_1,n)=2^n-2^{n/2}-1$, $N(d_2,n)=1$, and $N(d_3,n)=1$, respectively.

From the trivial degeneracy distribution of these examples of graphs, i.e., fully connected and bipartite fully connected, we see that the entropies approach trivial limits for large system sizes.
Namely, for the strong rule, using Eqs.~(\ref{resolution}) and (\ref{entropy}) for the output and degeneracy entropies, respectively, we see the in both types of graphs $H[y]$ and $H[d]$ both approach 0, since the distribution is dominated by a single degeneracy $d \cong 2^n$ with $N(d,n)=1$.
With the weak rule,the entropy $H[y]/n$ approaches $\ln 2 = 0.693…$ and $H[d]$ approaches 0.
In general, we expect that the entropies approach these limits was we increase the degree of the graphs generated by any model.
Interestingly, this effect is already quite visible in Tables \ref{Table2} and \ref{Table3}, when we compare the values of the entropy for different degrees within each class of graphs, even for degrees up to only 5.

%%%%%%

%%%%%%%%%%%%%%%%%%%%%%%%%%%%%%%%%%%%%%%%%%
\section{Discussion}

In Ref.~\cite{baxter2020complex} we introduced a simple filtering problem which produces a rich and complex distribution of output degeneracies. The input is a cyclic sequence of zeroes and ones (a ring), and the process outputs a one in any position where a particular short pattern occurs, and a zero otherwise. 
The tractability of the problem means that we are able to give the complete degeneracy distribution, for the set of all possible inputs, up to relatively large system sizes.

In this paper, we have extended this problem to consider general graphs. The input is a digit $1$ or $0$ assigned to each node of the graph, and the output for each node is $1$ if the state of the node and those of its immediate neighbours match a given filter pattern, and $0$ otherwise.
We demonstrate this process by calculating the full degeneracy distributions for various degree regular graphs with $30$ or more nodes, using two example filter patterns. The weak (WR) pattern registers a $1$ if the corresponding node has state $1$ and at least one of its neighbours has state $0$. The strong (SR) pattern only registers $1$ if the node is in state $1$ and all of its neighbours are in state $0$.
We found degeneracy distributions having similar form and features to those seen in the simpler problem of filtering on a ring.
We showed that three key features of the degeneracy distribution: the largest degeneracy $d_D(n)$, the number of distinct outputs $M(n)$ and the number of outputs having degeneracy one, $N(1,n)$ behave as $z_d^n$, $z_g^n$ and $z_a^n$, respectively, where the three numbers $z_d$, $z_g$ and $z_a$ take values from $1 $ to $2$ depending on the graph and the filter. We find precise values for these three numbers for all the graphs studied.

The two filter examples used give quite different results, and have different behaviour with respect to graph degree.
The key results are summarised by our main figure, Figure~\ref{numbers_vs_q}.
The weak rule filter, WR, is only weakly sensitive to the neighborhood of a node, and hence the structure of the graph. For large 
degree, it almost always produces an output matching the input. Thus the WR filter produces large values for the ouput entropy, called the resolution, and small values for the degeneracy entropy, the relevance.

The strong rule filter, SR, on the other hand, imposes a condition on all the neighbours of the node where the filter is applied. This produces a much larger relevance (which is a measure of the informativeness of the filtering process) in random graphs, but much lower resolution, as the number of unique outputs is restricted. The relevance is largest for the smallest graph degree not equal to two. Deterministically constructed graphs do not demonstrate the same peak in relevance, underlining the importance of this measure for detecting complexity. 
For larger degree, the condition becomes more restrictive, so the number of outputs is reduced. The resolution decreases with increasing $q$, but so does the relevance.
The reason that the $q=2$ graphs do not give the maximum relevance is that these graphs necessarily have a highly predictable structure. All nodes lie in one or at most a few rings. One may observe that the degeneracy distributions and corresponding statistics are very similar for all families of graphs studied when $q=2$.
The fact that results are largely determined by degree, indicates that it should be possible to write a mean field theory for the degeneracy distribution.

Similar complexity is observed in various complex systems, particularly with regard to information processing. In such systems, degeneracy distributions has been shown to be an important observation of the system.
The entropy of this distribution, called the relevance, was shown \cite{cubero2018minimally} to be the relevant measure of complexity, and we showed that our simple problem reproduces many of the important qualitative phenomena observed in such systems.
The filtering problem is therefore a highly tractable problem illuminating some of the key features of information processing in more complex systems. The extension of this problem to arbitrary graphs, makes the interactions between nodes more complex, and the analogy with the complex interactions of real complex systems more explicit.

%%%%

%%%%%%%%%%%%%%%%%%%%%%5
%%%%%%%%%%%%%%%%%%%%%%%%%
%%%%%%%%%%%%%%%%%%%%%%%%%
%%%%%%%%%%%%%%%%%%%%%%%%%%

%%%%%%%%%%%%%%%%%%%%%%%%%%%%%%%%%%%%%%%%%%
\section{Materials and Methods}

%%%%%%%%%%%%%%%%%%%

\subsection{Calculation of degeneracy distributions}
%\label{MM_exact-distribution}
\label{s3p}

The distributions shown in Figures~\ref{distribution_examples}-\ref{graphs_distributions2}, \ref{graphs_distributions4} and \ref{graphs_distributions5}, and the numbers presented in the Tables \ref{Table1}, \ref{Table2}, and \ref{Table3} and plotted in Figure~\ref{numbers_vs_q} were experimentally obtained by considering all $2^n$ configurations of the $n$ input binary variables $x_i$ individually.
For a specified filter, or rule, we obtain the output variables $y_i$ corresponding to each input.
From the frequency with which each output configuration appears, we build the degeneracy distribution.

For the sake of simplicity in the implementation of the computational experiments, we apply a basic indexing system to the output configurations.
We start by initializing an array with $2^n$ positions populated with zeros, representing the frequency of observation of each output.
Then, as we systematically run through all the possible inputs and calculate the corresponding outputs $\{y_i\}$, we increment by 1 the value in position $\sum_i y_i 2^i$ of the array, where $i=0,1,\dots,n-1$.
In the end of this process, each position of the array contains the frequency of its corresponding output.
This method is memory intensive, and in some cases uses much more memory than strictly necessary, since most of the positions of the frequency array will remain unchanged after initialization (corresponding to non-realizable, or unobserved, outputs).
It is relatively simple to develop methods that do not require so much memory, however they would necessarily require more CPU resources, and have a larger time complexity.
Notice that our method’s time complexity is linear with the number of input configurations $2^n$.
In the case of rings, a much more efficient algorithm may be used, as described in Ref.~\cite{baxter2020complex}.

%%%%%%%%%%%%%%%
%%%%%%%%%%
%%%%%%%%%%%%%%
%%%%%%%%%%%
%%%%%%%%%%%%%%%
%%%%%%%%%%
%%%%%%%%%%%%%%
%%%%%%%%%%%
%

\subsection{Asymptotics of the degeneracy distribution on rings}
\label{MM_asymptotics}

Here we show how the asymptotic behaviour of the degeneracy distribution may be obtained.
We focus on the particular family of filter patterns consisting of a chain of $1$s with a $0$ at each end. The shortest such pattern is $010$. Each member of this set may be indexed by the length of the filter, $w \geq 3$.
Each output consists of isolated ones separated by strings of zeroes of various lengths. 
The filter pattern length $w$ determines the minimum number of zeroes, $w-2$, between each one. 

For $w=3$, chains of  three or fewer zeroes in the output can only be produced in one way.
Thus outputs containing only such chains of zeroes have degeneracy $1$.
Possible such output sequences can be built up out 
of 
three kinds of building blocks, $01$, $001$, and $0001$, put together in a ring of length $n$. We can thus find the number of outputs of degeneracy $1$, ${\cal N}(1,n)$, by counting all possible ways of building a ring of length $n$ out of these blocks.
We can do this recursively. For every configuration of length $n-2$, we can obtain a valid configuration of length $n$ by inserting the block $01$ to the right, say, of a particular position $i$ in the ring. This gives all the configurations of length $n$ with the block $01$ to the right of $i$.
Doing the same with configurations of length $n-3$ and blocks $001$, we get all configurations with a block $001$ to the right of the block of $i$. 
Finally, repeating the procedure for configurations of length $n-4$ and blocks $0001$, gives all configurations with a block $0001$ to the right of the block of $i$.
Since every block must be $01$, $001$, or $0001$, the union of these three sets is the full set of configurations of degeneracy $1$ in rings of $n$ digits.
Thus, we can write
\begin{equation}
{\cal N}(1,n) = {\cal N}(1,n-2) + {\cal N}(1,n-3) + {\cal N}(1,n-4)
.
\label{eq90}
\end{equation}
Starting from the  first few values
 \begin{equation}
{\cal N}(1,1){=}0, \  {\cal N}(1,2){=}2, \ {\cal N}(1,3){=}3, \ {\cal N}(1,4){=}6
,
\label{eq105}
\end{equation}
we could build up the sequence and find ${\cal N}(1,n)$ for any $n$. However it is not necessary to iterate through all values of $n$. 

The explicit solution of this linear difference equation (\ref{eq90}) can be written in terms of the roots, $z_i$, of the characteristic equation $z^4 = z^2 + z +1$:
 \begin{equation}
{\cal N}(1,n) = {z_1}^n + {z_2}^n + {z_3}^n + {z_4}^n
,
\label{eq100}
\end{equation}
where the coefficients of the powers of the roots $z_i$, all equal to one, are found form the initial condition, Eq.~(\ref{eq105}).
The root $z_1 \equiv z_a = 1.46557...$ determines the large $n$ asymptotics of ${\cal N}(1,n)$.% , Eq.~(\ref{100}).

For $w\geq4$, it becomes possible for there to be chains of ones in the input that are shorter than that in the filter pattern. This means that only sequences of $w-2$ or $w-1$ zeroes in the output are not degenerate. 
Any sequence of $w$ or more zeroes in the output can be produced in more than one way. 
One may therefore extend an input of degeneracy $1$ only by inserting blocks of length $w-1$ and $w$. Hence the recursion for ${\cal N}(1,n)$ becomes
\begin{align}
{\cal N}(1,n) = {\cal N}(1,n-w+1) + {\cal N}(1,n-w)
.\label{N1w_recurs}
\end{align}
The corresponding characteristic equation is 
\begin{align}\label{chareq_N1w}
z^w = z+1.
\end{align}
For large $n$, then, 
\begin{align}
{\cal N}(1,n) \cong z_a^{n}
\end{align}
Where $z_a$ corresponds to the dominant solution  of Eq. (\ref{chareq_N1w}).

%%%
%%%

The total number of possible outputs may be derived in a similar way.
The presence of a $1$ at a given position in the output corresponds uniquely to $w$ fixed digits at the same position in the input. Any degeneracy therefore arises in the parts of the input corresponding to strings of zeroes in the output.
The total number of possible outputs, $M(n)$, is then the number of ways of arranging isolated ones in a chain of length $n$, subject to this constraint.
% %%
For every output of length $n-1$, we can create an output of length $n$ by inserting an additional $0$. The same is not true for the digit $1$, however. Any $1$ in the output must be accompanied by a sequence of $w-2$ zeroes. We can account for this condition precisely by inserting the sequence $10_{w-2}$ into any valid output of length $n-(w-1)$ in a position immediately following a sequence of $w-2$ zeroes (at least one such sequence must exist).
Thus $M (n) = M (n - 1) + M(n - w +1)$, with initial conditions $M(n=w) = 2$, $M(n<w) = 1$.
The elements of the sequence may be written in terms of the roots of the characteristic equation \cite{hoggatt1969fibonacci,graham1994concrete,koshy2019fibonacci}
\begin{align}
z^{w-1} = z^{w-2} + 1.
\end{align}
Then $z_g$ corresponds to the largest root of this equation. We list values for various filter lengths (as well as for some other filter patterns) in Table~\ref{Table1}.

The entire degeneracy distribution may be built up by considering chains of zeroes of different lengths in the output, and the number of different possible corresponding sections of the input.
Let an output with $m\geq 1$ ones contain $m$ strings of zeroes with lengths $\ell_1,\ell_2,...,\ell_m$. Then the degeneracy of this output equals 
\begin{equation}
d = \prod_{i=1}^{m}  \tilde{d}(\ell_i) 
. 
\label{41}
\end{equation}
Here $\tilde{d}(\ell)$ is the number of input strings of length $\ell$, having the first and last digits $0$, that  generate an output string of $\ell$ zeroes. 
This number plays an important role in our problem, similar to prime numbers in number theory, so we call the $\tilde{d}(\ell)$ {\em prime degeneracies}. 
Suppose that the output contains $\mu_\ell$ strings of zeroes of length $\ell$, $\ell=w-2,w-1,w,...$, where  
\begin{equation}
m + \sum_{\ell\geq  w-2} \ell \mu_\ell = n
.
\label{42}
\end{equation}
Then Eq.~(\ref{41}) may be rewritten 
\begin{equation}
d = \prod_{\ell\geq w-2} [\tilde{d}(\ell)]^{\,\mu_\ell}
%%.
\label{43}
\end{equation}
for $m\geq1$.

The prime degeneracies $\tilde{d}(\ell)$ 
can be obtained recursively by taking into account three points: 

(i) Relevant input configurations of length $\ell$ are obtained by inserting 
$0$ or 
$1$ into each relevant configuration of length $\ell-1$ between the first and second positions of the sequence. (Recall that the first and last positions of the input sequence are fixed to 0.) 

(ii) Input strings of length $\ell$ beginning and/or ending with $01_{w-2}0$ are irrelevant, and so they should be removed from the set generated at the previous step.  
These configurations can be obtained by inserting the $w-1$ digits $1_{w-2}0$ into each relevant input string of length $\ell-w+1$ between its first and second positions.

(iii) Finally, there exist input strings, compatible with the output string of $\ell$ zeroes, 
that cannot be obtained by inserting a single digit into relevant input strings of length $\ell-1$ between their first and second positions. These are the input strings of length $\ell$ beginning with $01_{w-1}0$ (i.e. a string of ones one digit longer than in the filter). These inputs can be obtained by inserting $1_{w-1}0$ into each relevant input string of length $\ell-w$ between their first and second positions.

Following these rules, the degeneracy of a string of $\ell$ zeroes at the output, prime degeneracy $\tilde{d}(\ell)$, can be written recursively as a linear difference equation: 
\begin{equation}
  \tilde{d}(\ell) = 2 \tilde{d}(\ell-1) - \tilde{d}(\ell-w+1) + \tilde{d}(\ell-w)
%%  ,
\label{eq10}
\end{equation} 
with the initial condition $\tilde{d}(1)=\tilde{d}(2)=1$,
$\tilde{d}(\ell)=2^{\ell-2}$ for $3 \leq \ell < w$ and $d(w) = 2^{w-2}-1$.
The solution of Eq.~(\ref{eq10}) may be explicitly expressed in terms of the complex roots %$z_1$, $z_2$, and $z_3$ 
of the characteristic equation 
\begin{equation}
z^w=2z^{w-1}-z+1
.
\label{12}
\end{equation} 

giving 
\begin{equation}
  \tilde{d}(\ell) = C_1 {z_1}^\ell + C_2 {z_2}^\ell  + C_3 {z_3}^\ell + ... + C_w {z_w}^\ell 
  .
\label{eq20}
\end{equation}
The largest real root of Eq.~(\ref{12}), $z_1$, say, dominates for large $\ell$, and we identify it as $z_d$:
%%%
\begin{equation}
\tilde{d}(\ell) \cong C_1z_d^\ell.%
\label{23}
\end{equation} 

The case of the periodic output of length $n$ with all digits 0 has to be considered separately. 
Consider one digit of the input, at an arbitrary position.
The number of input configurations where this digit is 0 and the resulting output has only zeroes is given by $\tilde{d}(n+1)$, because the periodicity of the input means that this digit 0 plays the role of both first and last digit of the configurations of a string of $n+1$ digits. If the digit is 1, then the number of input configurations equals $1+\sum_{i\neq w-2} i \tilde{d}(n-i)$, where the sum over $i$ accounts for the configurations where the digit is in a group of $i$ consecutive ones whose length is not $w-2$, plus one configuration with all input digits equal to 1.
Thus the degeneracy of the output with all zeroes is given by 
\begin{equation}
d_D(n) = 1 + \tilde{d}(n+1) + \sum_{i=1;i\neq w-2}^{n-1} i \tilde{d}(n-i)
, 
\label{eq21}
\end{equation}
which is the largest possible degeneracy of an output of a given length. 
Applying the recursion relation for prime degeneracies $\tilde{d}$, Eq.~(\ref{eq10}) 
 to the terms on the right-hand side of Eq.~(\ref{eq21}) 
we find that the largest degeneracy $d_D(n)$ satisfies the same difference equation as Eq.~(\ref{eq10}) 
though with different initial condition%,
\begin{equation}
d_D(n) = 2 d_D(n-1) - d_D(n-w+1) + d_D(n-w)
\end{equation} 
with the initial condition  $d_D(n) = 2^n$ for $n < w$, and $d_D(w) = 2^w-w$.
%%%
For large $n$, the solution is dominated by a single solution,
\begin{equation}
d_D(n) \cong z_d^n
. 
\label{27}
\end{equation}
\vspace{6pt} 

%%%%%%%%%%%%%%%%%%%%%%%%%%%%%%%%%%%%%%%%%%
%% optional
%\supplementary{The following are available online at \linksupplementary{s1}, Figure S1: title, Table S1: title, Video S1: title.}

% Only for the journal Methods and Protocols:
% If you wish to submit a video article, please do so with any other supplementary material.
% \supplementary{The following are available at \linksupplementary{s1}, Figure S1: title, Table S1: title, Video S1: title. A supporting video article is available at doi: link.}

%%%%%%%%%%%%%%%%%%%%%%%%%%%%%%%%%%%%%%%%%%
\authorcontributions{
All authors contributed equally substantially in all parts and aspects of the work.
%
%For research articles with several authors, a short paragraph specifying their individual contributions must be provided. The following statements should be used ``Conceptualization, X.X. and Y.Y.; methodology, X.X.; software, X.X.; validation, X.X., Y.Y. and Z.Z.; formal analysis, X.X.; investigation, X.X.; resources, X.X.; data curation, X.X.; writing--original draft preparation, X.X.; writing--review and editing, X.X.; visualization, X.X.; supervision, X.X.; project administration, X.X.; funding acquisition, Y.Y. All authors have read and agreed to the published version of the manuscript.'', please turn to the  \href{http://img.mdpi.org/data/contributor-role-instruction.pdf}{CRediT taxonomy} for the term explanation. Authorship must be limited to those who have contributed substantially to the work reported.
}
%%%%%%%%%%%%%%%%%%%%%%%%%%%%%%%%%%%%%%%%%%
\funding{
This work was developed within the scope of the project i3N, UIDB/50025/2020 
\& UIDP/50025/2020, financed by national funds through the FCT/MEC.
 This work was also supported by National Funds through FCT, I. P. Project No. IF/00726/2015. R. A. d. C. acknowledges the FCT Grant No. CEECIND/04697/2017.
%Please add: ``This research received no external funding'' or ``This research was funded by NAME OF FUNDER grant number XXX.'' and  and ``The APC was funded by XXX''. Check carefully that the details given are accurate and use the standard spelling of funding agency names at \url{https://search.crossref.org/funding}, any errors may affect your future funding.
}

%%%%%%%%%%%%%%%%%%%%%%%%%%%%%%%%%%%%%%%%%%
%\acknowledgments{In this section you can acknowledge any support given which is not covered by the author contribution or funding sections. This may include administrative and technical support, or donations in kind (e.g., materials used for experiments).}

%%%%%%%%%%%%%%%%%%%%%%%%%%%%%%%%%%%%%%%%%%
\conflictsofinterest{The authors declare no conflict of interest.} 

%%%%%%%%%%%%%%%%%%%%%%%%%%%%%%%%%%%%%%%%%%
%% optional
%\abbreviations{The following abbreviations are used in this manuscript:\\
%
%\noindent 
%\begin{tabular}{@{}ll}
%MDPI & Multidisciplinary Digital Publishing Institute\\
%DOAJ & Directory of open access journals\\
%TLA & Three letter acronym\\
%LD & linear dichroism
%\end{tabular}}

%%%%%%%%%%%%%%%%%%%%%%%%%%%%%%%%%%%%%%%%%%
%% optional
% \appendixtitles{no} % Leave argument "no" if all appendix headings stay EMPTY (then no dot is printed after "Appendix A"). If the appendix sections contain a heading then change the argument to "yes".
 \appendix
% \section{}
% \unskip
% \subsection{}
% The appendix is an optional section that can contain details and data supplemental to the main text. For example, explanations of experimental details that would disrupt the flow of the main text, but nonetheless remain crucial to understanding and reproducing the research shown; figures of replicates for experiments of which representative data is shown in the main text can be added here if brief, or as Supplementary data. Mathematical proofs of results not central to the paper can be added as an appendix.

% \section{}
% All appendix sections must be cited in the main text. In the appendixes, Figures, Tables, etc. should be labeled starting with `A', e.g., Figure A1, Figure A2, etc. 

\section{Further results for the weak rule filter}\label{APP_WRresults}

Here we plot degeneracy distributions, cumulative distributions, and tabulate measures for the weak rule filter, WR, for comparison with those given for the strong rule, SR, in the main body of the text above.
 %%%=================================================================
 %%%=================================================================
 \begin{figure}[H]
 \centering
 \includegraphics[width=0.69\textwidth]{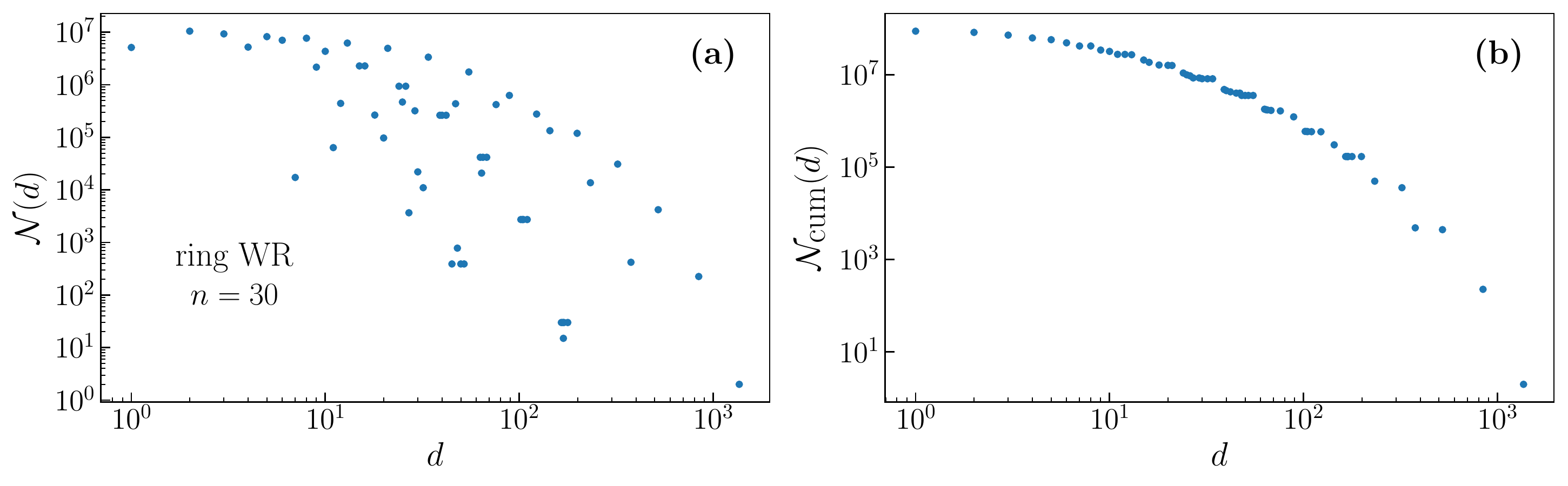}
 \includegraphics[width=0.69\textwidth]{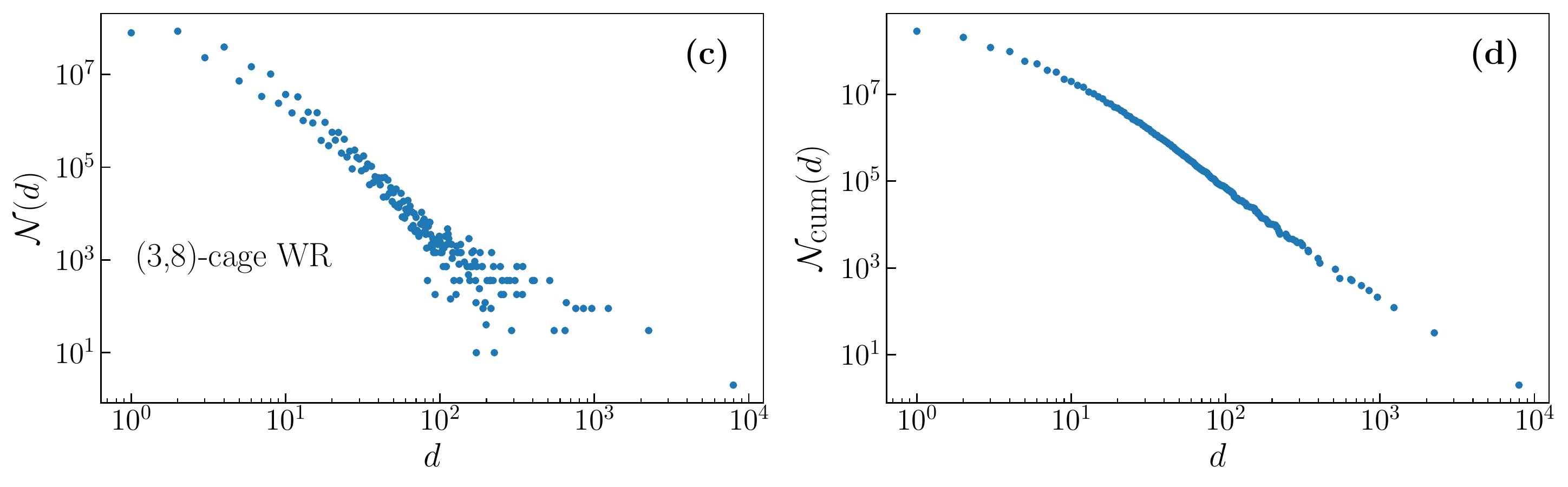}
 \includegraphics[width=0.69\textwidth]{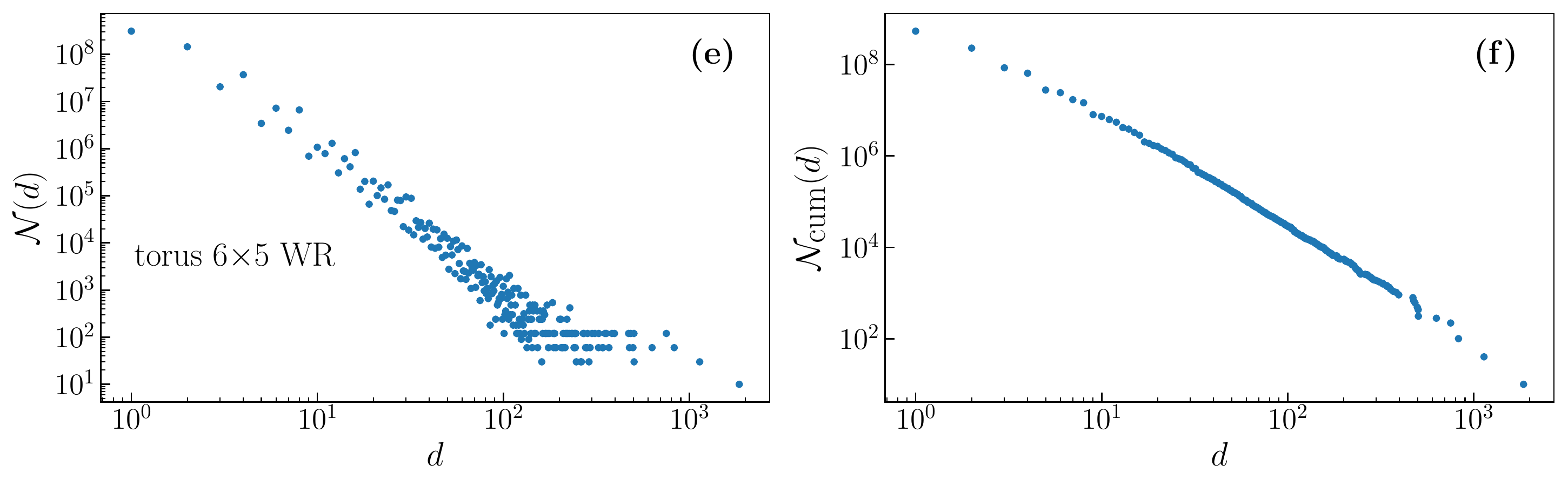}
 \caption{
   Degeneracy distributions (left) and cumulative degeneracy distributions (right) for outputs of the WR filter 
   on selected deterministic graphs of degree 2 (a,b) 3 (c,d) and 4 (e,f).
 }\label{graphs_distributions4}
 \end{figure}   
 %%%=================================================================
 %%%=================================================================

 %%%=================================================================
 %%%=================================================================
 \begin{figure}[H]
 \centering
 \includegraphics[width=0.69\textwidth]{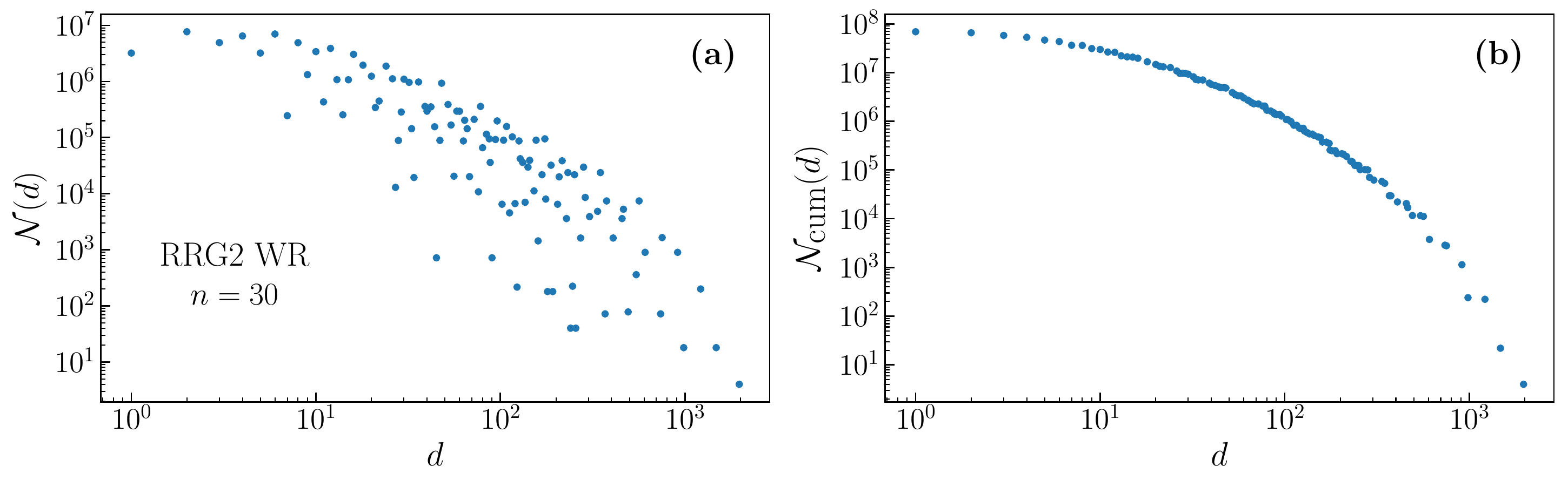}
 \includegraphics[width=0.69\textwidth]{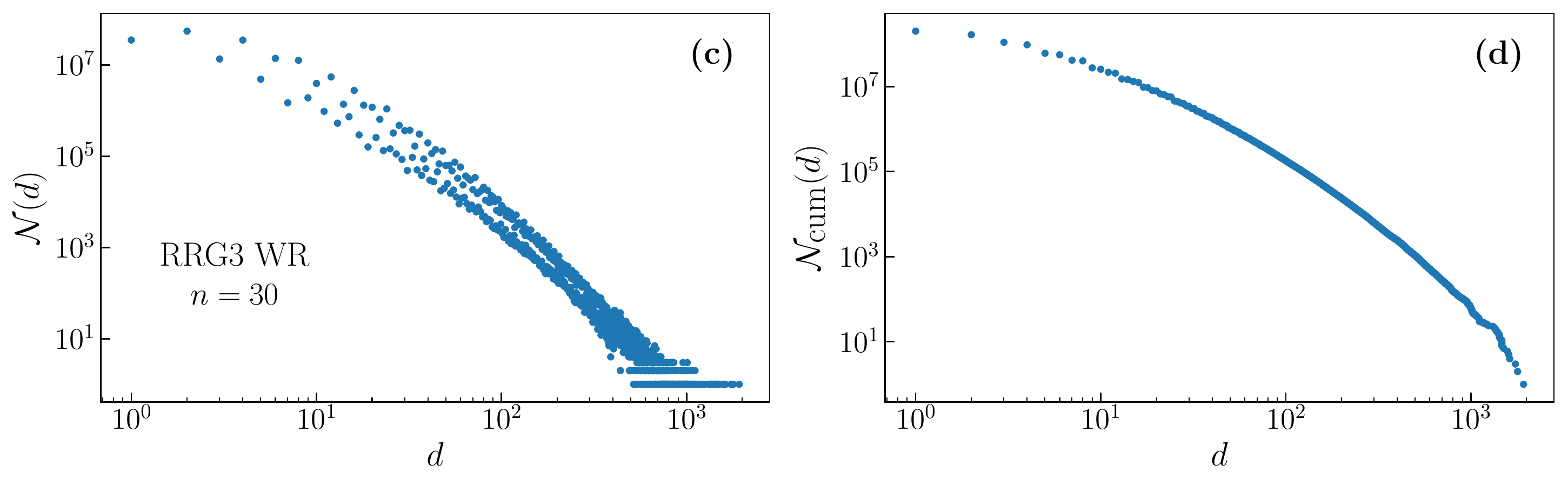}
 \includegraphics[width=0.69\textwidth]{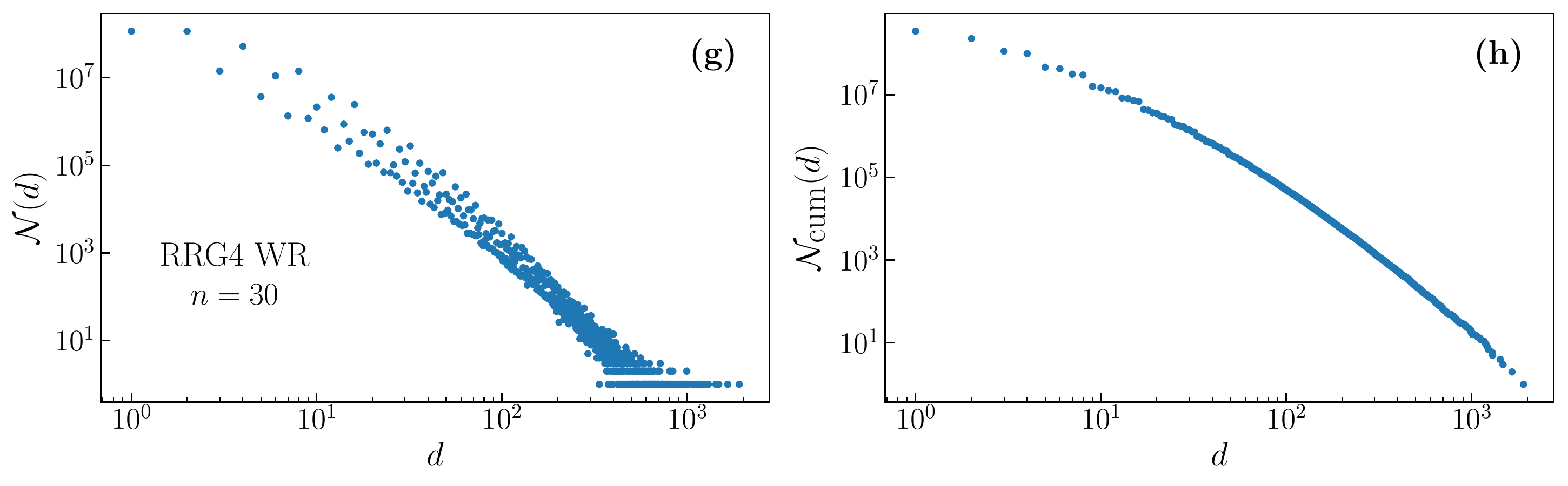}
 \caption{
  Degeneracy distributions and cumulative degeneracy distributions for outputs of the SR filter on random regular graphs of degree 2 (a,b) 3 (c,d) and 4 (e,f).
 }\label{graphs_distributions5}
 \end{figure}   
 %%%=================================================================
 %%%=================================================================

%%%%%%%%%%%%%%%%%%%%%%%%%%%%%%%%%%%%%%%%%%%
%%%%%%%%%%%%%%%%%%%%%%%%%%%%%%%%%%%%%%%%%%%
%%%%%%%%%%%%%%%%%%%%%%%%%%%%%%%%%%%%%%%%%%%
\begin{table}[H]
\caption{
Important values for the degeneracy distribution resulting from applying the weak rule (WR) filter to various graphs.
The numbers $\sqrt[{\scriptstyle n}]{M(n)}$,  $\sqrt[{\scriptstyle n}]{d_D(n)}$ and  $ \sqrt[{\scriptstyle n}]{N(1,n)}$ approximate $z_g$,  $z_d$ and $z_a$ respectively. 
We also give the relevance per node $H[d]/n$ and the resolution per node $H[y]/n$.
Numbers for RRG($q$) and SW($q$) were obtained by averaging over 10 random realizations.
} \label{Table3}
\centering
%\tb newline within a bracket \ts new line of same size not bracketed, \tn new si<e
\begin{tabular}{l c l l c c c p{0.1\textwidth}}%c c c c c c p{0.1\textwidth}}
\toprule    
%graph         nodes(n)     H[d] (nats)             H[y] (nats) & # degeneracies[D(n)] & max degen[d_D(n)]    # outputs[M(n)]    # outputs of d=1[N(1,n)]
%
graph         & $n$  &  $\sqrt[{\scriptstyle n}]{M(n)}$   & $\sqrt[{\scriptstyle n}]{d_D(n)}$    & $\sqrt[{\scriptstyle n}]{N(1,n)}$ &   $H[d]/n$     &   $H[y]/n$  \\
\midrule
Apollonian 2  & 7&  	1.95461	 & 	1.21901	 & 	1.91660	 & 	0.11519	 & 	0.66045	   \\
Apollonian 3  & 16 &  1.94788	 & 	1.43435	 & 	1.91189	 & 	0.09594	 & 	0.64711	   \\
\midrule
(3,5)-cage & 10 & 1.91202	 & 	1.21481	 & 	1.84295	 & 	0.13705	 & 	0.62974	   \\
(3,6)-cage &  14 &  	1.91394	 & 	1.34590	 & 	1.83757	 & 	0.12271	 & 	0.63149	   \\
(3,7)-cage &  24 &  	1.91348	 & 	1.25055	 & 	1.83511	 & 	0.11462	 & 	0.63259	   \\
(3,8)-cage & 30 &  	1.91330	 & 	1.34897	 & 	1.83337	 & 	0.10559	 & 	0.63275	   \\
(4,5)-cage & 19 &  	1.95248	 & 	1.21101	 & 	1.91027	 & 	0.08217	 & 	0.65878	   \\
(4,6)-cage & 26 &  	1.95322	 & 	1.37995	 & 	1.91085	 & 	0.07188	 & 	0.65902	   \\
(5,5)-cage 1 & 30 &  	1.97461	 & 	1.16392	 & 	1.95220	 & 	0.04494	 & 	0.67453	   \\
(5,5)-cage 2 & 30 &  	1.97461	 & 	1.18854	 & 	1.95219	 & 	0.04495	 & 	0.67453	   \\
(5,5)-cage 3 & 30 &  	1.97461	 & 	1.21540	 & 	1.95220	 & 	0.04496	 & 	0.67453	   \\
(5,5)-cage 4 & 30 &  	1.97461	 & 	1.17585	 & 	1.95220	 & 	0.04495	 & 	0.67453	   \\
\midrule
torus $3{\times}3$ & 9 &  	1.95698	 & 	1.16653	 & 	1.92324	 & 	0.10088	 & 	0.66192	   \\
torus $4{\times}4$ &  16 &  	1.95546	 & 	1.38485	 & 	1.92191	 & 	0.09024	 & 	0.65777	   \\
torus $5{\times}5$ &  25 &  	1.95475	 & 	1.21993	 & 	1.91904	 & 	0.08076	 & 	0.65828	   \\
torus $6{\times}5$ &  30 &  	1.95475	 & 	1.28517	 & 	1.91898	 & 	0.07568	 & 	0.65831	   \\
torus $10{\times}3$ &  30 &  	1.95626	 & 	1.22522	 & 	1.91924	 & 	0.07072	 & 	0.66127	   \\
torus $8{\times}4$ &  32 &  	1.95510	 & 	1.38392	 & 	1.91932	 & 	0.07475	 & 	0.65813	   \\
torus $6{\times}6$ & 36 &  	1.95475	 & 	1.38400	 & 	1.91883	 & 	0.07034	 & 	0.65833	   \\
\midrule
SWB(3) & 10 &  	1.91492	 & 	1.35588	 & 	1.85212	 & 	0.14568	 & 	0.62873	   \\
SWB(3) & 20 &  	1.91523	 & 	1.30100	 & 	1.84849	 & 	0.12731	 & 	0.62956	   \\
SWB(3) & 30 &  	1.91523	 & 	1.35620	 & 	1.84851	 & 	0.11281	 & 	0.62956	   \\
SWB(4) & 12 &  	1.95603	 & 	1.17605	 & 	1.91983	 & 	0.09152	 & 	0.66087	   \\
SWB(4) & 21 &  	1.95626	 & 	1.21231	 & 	1.91923	 & 	0.08079	 & 	0.66127	   \\
SWB(4) & 30 &  	1.95626	 & 	1.20790	 & 	1.91924	 & 	0.07071	 & 	0.66127	   \\
SWB(5) & 12 &  	1.97929	 & 	1.17605	 & 	1.96131	 & 	0.05848	 & 	0.67803	   \\
SWB(5) & 20 &  	1.97927	 & 	1.16442	 & 	1.96025	 & 	0.04881	 & 	0.67840	   \\
SWB(5) & 32 &  	1.97927	 & 	1.14893	 & 	1.96021	 & 	0.04144	 & 	0.67842	   \\
\midrule
RRG(2) & 10 &  	1.76075	 & 	1.33214	 & 	1.62827	 & 	0.16029	 & 	0.53010	   \\
RRG(2) & 20 &  	1.81141	 & 	1.29593	 & 	1.65746	 & 	0.14345	 & 	0.56553	   \\
RRG(2) & 30 &  	1.83115	 & 	1.27447	 & 	1.67090	 & 	0.11728	 & 	0.58305	   \\
RRG(3) & 10 &  	1.86350	 & 	1.31014	 & 	1.73760	 & 	0.18497	 & 	0.59914	   \\
RRG(3) & 20 &  	1.88987	 & 	1.28690	 & 	1.78608	 & 	0.13766	 & 	0.61721	   \\
RRG(3) & 30 &  	1.89895	 & 	1.27941	 & 	1.80483	 & 	0.11537	 & 	0.62310	   \\
RRG(4) & 10 &  	1.92754	 & 	1.24293	 & 	1.86647	 & 	0.13602	 & 	0.64106	   \\
RRG(4) & 20 &  	1.93917	 & 	1.25076	 & 	1.88272	 & 	0.09610	 & 	0.65019	   \\
RRG(4) & 30 &  	1.94507	 & 	1.25247	 & 	1.89654	 & 	0.07819	 & 	0.65347	   \\
RRG(5) & 10 &  	1.93764	 & 	1.26982	 & 	1.88340	 & 	0.12318	 & 	0.64808	   \\
RRG(5) & 20 &  	1.95952	 & 	1.23790	 & 	1.92369	 & 	0.07520	 & 	0.66371	   \\
RRG(5) & 30 &  	1.96616	 & 	1.22872	 & 	1.93641	 & 	0.05720	 & 	0.66839	   \\
\midrule
SW(3) & 10 &  	1.90692	 & 	1.25638	 & 	1.82855	 & 	0.15381	 & 	0.62737	   \\
SW(3) & 20 &  	1.90025	 & 	1.27244	 & 	1.80865	 & 	0.13264	 & 	0.62404	   \\
SW(3) & 30 &  	1.91093	 & 	1.27260	 & 	1.82941	 & 	0.10905	 & 	0.63105	   \\
SW(4) & 10 &  	1.91972	 & 	1.26865	 & 	1.85154	 & 	0.14500	 & 	0.63556	   \\
SW(4) & 20 &  	1.93992	 & 	1.25544	 & 	1.88853	 & 	0.09904	 & 	0.64926	   \\
SW(4) & 30 &  	1.94584	 & 	1.26608	 & 	1.89776	 & 	0.07874	 & 	0.65403	   \\
SW(5) & 10 &  	1.94942	 & 	1.23419	 & 	1.90650	 & 	0.11008	 & 	0.65650	   \\
SW(5) & 20 &  	1.96150	 & 	1.23542	 & 	1.92709	 & 	0.07365	 & 	0.66525	   \\
SW(5) & 30 &  	1.96889	 & 	1.23140	 & 	1.94184	 & 	0.05471	 & 	0.67024	   \\
 \bottomrule
 \end{tabular}
\end{table}
%%%%%%%%%%%%%%%%%%%%%%%%%%%%%%%%%%%%%%%%%%%
%%%%%%%%%%%%%%%%%%%%%%%%%%%%%%%%%%%%%%%%%%%
%%%%%%%%%%%%%%%%%%%%%%%%%%%%%%%%%%%%%%%%%%%

%%%%%%%%%%%%%%%%%%%%%%%%%%%%%%%%%%%%%%%%%%
\reftitle{References}

% Please provide either the correct journal abbreviation (e.g. according to the “List of Title Word Abbreviations” http://www.issn.org/services/online-services/access-to-the-ltwa/) or the full name of the journal.
% Citations and References in Supplementary files are permitted provided that they also appear in the reference list here. 

%=====================================
% References, variant A: external bibliography
%=====================================
\externalbibliography{yes}
\bibliography{degeneracy_distribution}

%=====================================
% References, variant B: internal bibliography
%=====================================
% \begin{thebibliography}{999}
% % Reference 1
% \bibitem[Author1(year)]{ref-journal}
% Author1, T. The title of the cited article. {\em Journal Abbreviation} {\bf 2008}, {\em 10}, 142--149.
% % Reference 2
% \bibitem[Author2(year)]{ref-book}
% Author2, L. The title of the cited contribution. In {\em The Book Title}; Editor1, F., Editor2, A., Eds.; Publishing House: City, Country, 2007; pp. 32--58.
% \end{thebibliography}

% The following MDPI journals use author-date citation: Arts, Econometrics, Economies, Genealogy, Humanities, IJFS, JRFM, Laws, Religions, Risks, Social Sciences. For those journals, please follow the formatting guidelines on http://www.mdpi.com/authors/references
% To cite two works by the same author: \citeauthor{ref-journal-1a} (\citeyear{ref-journal-1a}, \citeyear{ref-journal-1b}). This produces: Whittaker (1967, 1975)
% To cite two works by the same author with specific pages: \citeauthor{ref-journal-3a} (\citeyear{ref-journal-3a}, p. 328; \citeyear{ref-journal-3b}, p.475). This produces: Wong (1999, p. 328; 2000, p. 475)

%% for journal Sci
%\reviewreports{\\
%Reviewer 1 comments and authors’ response\\
%Reviewer 2 comments and authors’ response\\
%Reviewer 3 comments and authors’ response
%}

%%%%%%%%%%%%%%%%%%%%%%%%%%%%%%%%%%%%%%%%%%
\end{document}